\documentclass{emulateapj}
\shorttitle{Eccentric KOI-1474.01}
\shortauthors{Dawson et al.}

\def\kep{\emph{Kepler\ }}
\def\prob{{\rm prob}}
\def\msun{M_\odot}
\def\mstar{M_\star}
\def\rstar{R_\star}
\def\taustar{\tau_\star}
\def\rhostar{\rho_\star}
\def\acirc{a_{\rm circ}}
\def\afinal{a_{\rm final}}
\def\pfinal{P_{\rm final}}                                                                                                                                                                                                           
\def\prot{P_{\rm rot}}
\def\teff{T_{\rm eff}}
\def\fe{{\rm [Fe/H]}}
\def\vsini{v_{\rm rot} \sin i_s}
\begin{document}
\title{The Photoeccentric Effect and Proto-Hot Jupiters II. KOI-1474.01, a candidate eccentric planet perturbed by an unseen companion}
\slugcomment{Received 2012 July 7; accepted 2012 October 13}
\author{Rebekah I. Dawson\altaffilmark{1,2}}
\altaffiltext{1}{Harvard-Smithsonian Center for Astrophysics, 60 Garden St, MS-10, Cambridge, MA 02138}
\altaffiltext{2}{{\tt  rdawson@cfa.harvard.edu}}
\author{John Asher Johnson\altaffilmark{3,4}}
\altaffiltext{3}{Department of Astronomy, California Institute of Technology, 1200 East California Boulevard, MC 249-17, Pasadena, CA 91125, USA}
\altaffiltext{4}{NASA Exoplanet Science Institute (NExScI), CIT Mail Code 100-22, 770 South Wilson Avenue, Pasadena, CA 91125}
\author{Timothy D. Morton\altaffilmark{3}}
\author{Justin R. Crepp\altaffilmark{5}}
\altaffiltext{5}{Department of Physics, University of Notre Dame, 225 Nieuwland Science Hall, Notre Dame, IN 46556}
\author{Daniel C. Fabrycky\altaffilmark{6}}
\altaffiltext{6}{Department of Astronomy and Astrophysics, University of Chicago, 5640 S. Ellis Ave, Chicago, IL 95064}
\author{Ruth A. Murray-Clay\altaffilmark{1}}
\author{Andrew W. Howard\altaffilmark{7}}
\altaffiltext{7}{Institute for Astronomy, University of Hawaii, 2680 Woodlawn Drive, Honolulu, HI 96822-1839 }

\begin{abstract}
The exoplanets known as hot Jupiters---Jupiter-sized planets with periods less than 10 days---likely are relics of dynamical processes that shape all planetary system architectures. Socrates et al. (2012) argued that high eccentricity migration (HEM) mechanisms proposed for situating these close-in planets should produce an observable population of highly eccentric proto-hot Jupiters that have not yet tidally circularized. HEM should also create failed-hot Jupiters, with periapses just beyond the influence of fast circularization. Using the technique we previously presented for measuring eccentricities from photometry (the ``photoeccentric effect''), we are distilling a collection of eccentric proto- and failed-hot Jupiters from the \kep Objects of Interest (KOI). Here we present the first, KOI-1474.01, which has a long orbital period (69.7340 days) and a large eccentricity $e = 0.81^{+0.10}_{-0.07}$, skirting the proto-hot Jupiter boundary. Combining \kep photometry, ground-based spectroscopy, and stellar evolution models, we characterize host KOI-1474 as a rapidly-rotating F-star. Statistical arguments reveal that the transiting candidate has a low false-positive probability of 3.1\%. KOI-1474.01 also exhibits transit timing variations of order an hour. We explore characteristics of the third-body perturber, which is possibly the ``smoking-gun'' cause of KOI-1474.01's large eccentricity. Using the host-star's rotation period, radius, and projected rotational velocity, we find KOI-1474.01's orbit is marginally consistent with aligned with the stellar spin axis, although a reanalysis is warranted with future additional data. Finally, we discuss how the number and existence of proto-hot Jupiters will not only demonstrate that hot Jupiters migrate via HEM, but also shed light on the typical timescale for the mechanism.
\end{abstract}

\section{Introduction}

The start of the exoplanet era brought with it the discovery of an exotic new class of planets: Jupiter-sized bodies with short-period orbits ($P \lesssim 10$~days), commonly known as hot Jupiters \citep{1995M,1997M}. Most theories require formation of Jupiter-sized planets at or beyond the so-called ``snow line,'' located at roughly a few AU,\footnote{\citet{2008KB} and \citet{2008KK} explore in detail the location of the ice line for different stellar and disk parameters.} and debate the mechanisms through which they ``migrated'' inward to achieve such small semimajor axes. The leading theories fall into two categories: smooth migration through the proto planetary disk \citep[e.g.][]{1980G,1997W,2005AM,2008I,2011BK}, or what \citet{2012SK} (hereafter S12) term high eccentricity migration (HEM), in which the planet is perturbed by another body onto an inclined and eccentric orbit that subsequently circularizes through tidal dissipation \citep[e.g][]{2003W,2006F,2007FT,2011NF,2011WL}

From the present-day orbits of exoplanets we can potentially distinguish between mechanisms proposed to shape the architectures of planetary systems during the early period of dynamical upheaval. In this spirit, \citet{2011MJa} used the distribution of stellar obliquities to estimate the fraction of hot Jupiters on misaligned orbits and to distinguish between two specific migration mechanisms \citep[see also][]{2009F,2010TC,2010WF}; \citet{2012NF} recently applied a similar technique to estimate the relative contributions of two different mechanisms. However, deducing dynamical histories from the \emph{eccentricity} distribution of exoplanets poses a challenge because most hot Jupiters have already undergone tidal circularization and ``cold'' Jupiters at larger orbital distances may have formed in situ. Furthermore, type-II (gap-opening) migration may either excite or damp a planet's eccentricity through resonance torques \citep{2003G,2004S}. Finally, \citet{2011GR} find evidence that some hot Jupiters may have undergone disk migration either prior to or following scattering. In the latter case, disk migration may have damped their eccentricities. The eccentricity distribution is potentially shaped by a combination of HEM, tidal circularization, and planet-disk interactions.

Motivated by the HEM mechanisms proposed by \citet{2003W} and others, S12 proposed an observational test for HEM. As an alternative to modeling the \emph{distribution} of eccentricities, they suggested that we look for the \emph{individual} highly eccentric, long-period progenitors of hot Jupiters caught of the act of tidal circularization. S12 identified HD\,80606\,b as one such progenitor, which was originally discovered by radial velocity (RV) measurements of its host star's reflex motion \citep{2001N} and later found to transit along an orbit that is misaligned with respect to its host star's spin axis \citep{2009M,2009WH}. From statistical arguments S12 predicted that if HEM produces the majority of hot Jupiters, the \kep Mission should detect several ``super-eccentric'' Jupiters with orbital periods less than 93 days and eccentricities in excess of 0.9. A couple of these planets should be proto-hot Jupiters, with post-circularization semimajor axes in the region where all hot Jupiters have circularized (i.e. $P < 5$ days). Several more eccentric planets should have final periods above 5 days, in the region where not all hot Jupiters have circularized; these planets may be ``failed-hot Jupiters'' that will never circularize over their host stars' lifetimes. A failed-hot Jupiter may have either halted at its post-HEM location due to the tidal circularization timescale exceeding the age of the system, or undergone some tidal circularization but subsequently stalled after a perturber in the system raised its periapse. S12's prediction is supported by the existence of super-eccentric eclipsing binaries in the \kep sample, which are also thought to have been created by HEM mechanisms \citep{2012D}.

To test the HEM hypothesis we are ``distilling'' eccentric, Jupiter-sized planets  from the sample of announced \kep candidates using the publicly released \kep light curves \citep{2011BKB,2012B}. We described the distillation process and our technique for measuring eccentricities from transit light curves based on the ``photoeccentric effect'' in \citet{2012DJ}, hereafter Paper I. In summary, eccentric Jupiters are readily identified by their short ingress/egress/total transit durations \citep{2007B,2008F,2008B,2012P,2012KC}. A Markov-Chain Monte Carlo (MCMC) exploration of the posterior distributions of the transit parameters, together with a loose prior imposed on the stellar density, naturally accounts for the eccentricity-dependent transit probability and marginalizes over the periapse angle, yielding a tight measurement of a large orbital eccentricity (Paper I).

Here we present the first eccentric, Jupiter-sized candidate from the \kep sample: \kep Object of Interest (KOI) number 1474.01. We find that this eccentric candidate also has large transit-timing variations (TTVs). In fact, the TTVs are so large that they were likely missed by the automatic TTV-detection algorithms, as they were not listed in a recent cataloging of TTV candidates \citep{2012FR,2012S}. \citet{2011B} recently deduced the presence and planetary nature of the non-transiting Kepler-19c from the TTVs it caused in the transiting planet Kepler-19b, demonstrating the viability of detecting non-transiting planets through TTVs. More recently, \citet{2012N} characterized a Saturn-mass non-transiting planet using this technique. Thus the TTVs of 1474.01 may place constraints on the nature of an additional, unseen companion, thereby elucidating the dynamical history of the system.

In \S \ref{sec:lightcurve}, we present the light curve of KOI-1474.01. In \S \ref{sec:star}, we characterize the host-star KOI-1474 using \kep photometry, ground-based spectroscopy, and stellar evolution models. In \S \ref{sec:fpp}, we estimate the candidate's false positive probability (FPP) to be 3.1\%. In \S \ref{sec:ecc}, we measure KOI-1474.01's large eccentricity, investigate its TTVs and the perturbing third body that causes them, and measure the projected alignment of the transiting planet's orbit with the host star's spin axis. In \S \ref{sec:protofail}, we place KOI-1474.01 in the context of known hot Jupiters, proto-hot Jupiters, and failed-hot Jupiters, and explore whether KOI-1474.01 is a failed-hot Jupiter that will retain its current orbit or a proto-hot Jupiter that will eventually circularize at a distance close to the host star. We conclude in \S \ref{sec:discuss} by discussing the implications for planetary system formation models and suggesting directions for future follow up of highly eccentric planets in the \kep sample.
\begin{figure}[h]
\begin{centering}
\includegraphics{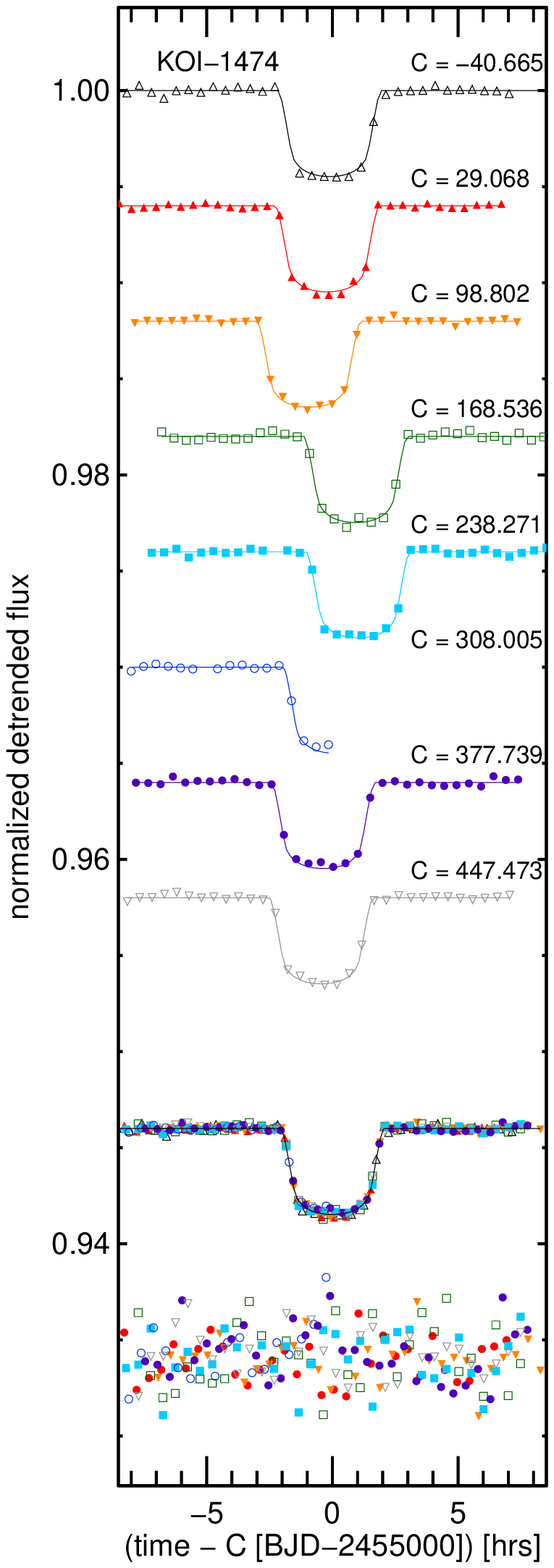}
\end{centering}
\end{figure}
\begin{figure}[t]
\begin{centering} 
\caption{Detrended light curves, color-coded by transit epoch, spaced with arbitrary vertical offsets. The top eight light curves are phased based on a constant, linear ephemeris (Table \ref{tab:planparams}, column 3), revealing the large TTVs. Each light curve is labeled `C' with its best-fit mid-transit time (Table \ref{tab:planparams}, column 3). In the second-from-the-bottom compilation, each light curve is shifted to have an individual best-fitting mid-transit time at t=0. The bottom points are the residuals multiplied by 10. Solid lines: best-fitting eccentric model (Table \ref{tab:planparams}, column 3). \label{fig:lightcurves}}
\end{centering}
\end{figure}

\section{KOI-1474.01: an interesting object of interest}
\label{sec:lightcurve}

KOI-1474.01 was identified by \citet{2011BKB} as an 11.3 R$_\oplus$ candidate that transits its 1.23 $\msun$, $6498$~K host star every 69.74538 days \citep{2010B}. With a \kep\ bandpass magnitude $K_P= 13.005$, the star is one of the brighter candidates in the \kep\ sample, making it amenable to follow up by Doppler spectroscopy. We retrieved the Q0-Q6 data from the Multimission Archive at the Space Telescope Science Institute (MAST) and detrended the light curve using {\tt AutoKep} \citep{2011G}. We identified eight transits (Figure \ref{fig:lightcurves}), which together reveal three notable properties:

\begin{enumerate}
\item When folded at a constant period, the transits are not coincident in phase. Indeed, some fall early or late by a noticeable fraction of a transit duration.
\item The transit durations are short for a planet with such a long orbital period (the total transit duration, from first to last contact, is $T_{14}$=2.92 hours, or 0.17$\%$ of the 69.74538 day orbital period). Yet instead of the V shape characteristic of a large impact parameter, the transit light curves feature short ingresses and egresses---  corresponding to a planet moving at 3 times the circular Keplerian velocity [based on the \kep Input Catalog (KIC) stellar parameters]---and a nearly flat bottom, implying that either the planet has a large eccentricity or orbits a very dense star (see Paper I). The candidate's reported $a/\rstar = 129.0525 \pm 0.0014$ \citep{2011BKB} corresponds to a stellar density of $6\rho_\odot$, which is inconsistent with main-sequence stellar evolution for all stars but late M-dwarfs. This implausibly high density derived from a circular orbital fit to the light curve implies that the planet has an eccentric orbit and is transiting near periapse (e.g. Figure 1 of Paper I).
\item The in-transit data feature structures that may be caused by star spot crossings (e.g. the bump in the purple, solid circle light curve marked C=377.739 in Figure \ref{fig:lightcurves}). The ratio of scatter inside of transits to that outside of transits is about 1.2. If the star exhibits photometric variability due to the rotation of its spot pattern, we may be able to measure the stellar rotation period and combine it with other stellar parameters to constrain the line-of-sight component of the system's spin-orbit configuration \citep[e.g.][]{2011S,2011NFF,2011DC}. If the star's surface temperature were greater than or equal to the KIC estimate of 6498 K \citep{2010B}, we might expect the star to lack a convective envelope \citep{2001P} and star spots. Therefore the star may be significantly cooler than this estimate.\footnote{However, \citet{2012HS} recently found photometric variability due to star spots for several hot stars, including KOI-1464, which has a surface temperature of $6578 \pm 70$ K, so the signatures of star spots we notice are not necessarily inconsistent with KOI-1474's KIC temperature.}
\end{enumerate}

The light curve implies that the transiting candidate KOI-1474.01 may be an eccentric planet experiencing perturbations from an unseen companion and with a measurable spin-orbit alignment, an ideal testbed for theories of planetary migration. However, in order to validate and characterize the candidate, first we must pin down the stellar properties and assess the probability that the apparent planetary signal is a false positive.

\section{Host KOI-1474, a rapidly-rotating F star}
\label{sec:star}

The stellar properties of KOI-1474 are essential for validating and characterizing the transiting candidate, but the parameters in the KIC are based on broadband photometry and may be systematically in error, as noted by \citet{2011BL}.  Here we use a combination of spectroscopy (\S \ref{subsec:spec}), photometry (\S \ref{subsec:rot}), and stellar evolution models (\S \ref{subsec:stellar}) to characterize host star KOI-1474.

	\subsection{Stellar temperature, metallicity, and surface gravity from spectroscopy}
\label{subsec:spec}

We obtained two high signal-to-noise, high resolution spectra for KOI-1474 using the HIgh Resolution Echelle Spectrometer (HIRES) on the Keck I Telescope \citep{1994V}. The spectra were observed using the standard setup of the California Planet Survey, with the red cross disperser and the $0\farcs86$ C2 decker, but with the iodine cell out of the light path \citep{2012J}. The first observation was made with an exposure time of 270 seconds, resulting in a signal-to-noise ratio (SNR) of $\approx45$ at 6000$~\AA$; the second exposure was 1200 seconds long, resulting in a SNR $\approx 90$. 

As described in Paper I, we use {\tt SpecMatch} to compare the two spectra to the California Planet Survey's vast library of spectra for stars with parameters from \emph{Spectroscopy Made Easy} \citep[SME;][]{1996V,2005V}. The closest-matching spectrum is that of HD\,3861. In order to match KOI-1474 to this relatively slowly rotating F dwarf, we must rotationally broaden the spectrum of HD\,3861. The total line broadening for KOI-1474, $\vsini = 13.6 \pm$ 0.5 km/s, is a combination of the HIRES instrumental profile, rotational broadening, and broadening due to turbulence (macroturbulence being the dominant term, rather than microturbulence: \citealt{2005V}). We assume that KOI-1474 has the same macroturbulent broadening and instrumental profile as HD\,3861. Then we apply additional rotational broadening to HD\,3861 using MORPH \citep{2006J} to match the spectra of KOI-1474 using the rotational broadening kernel described by \citet{2008G}. The $\vsini$ for KOI-1474 is $$\vsini = \sqrt{(v_{\rm rot} \sin i_{s})_{\rm HD 3861}^2+(v_{\rm rot} \sin i_{s})_{\rm broad}^2}$$ where $ (v_{\rm rot} \sin i_{s})_{\rm HD 3861} = 2.67$ km/s is the known $\vsini$ of HD 3861 \citep{2005V} and $(v_{\rm rot} \sin i_{s})_{\rm broad} = 13.3$ km/s is the additional rotational broadening applied to the HD\,3861 spectrum to match the lines of KOI-1474. See \citet{2011A}, \S 3.1 for a discussion and demonstration of this technique for measuring $\vsini$.
	
Next, from a weighted average of the properties of HD 3861 and the other best match spectra, we measure an effective temperature $\teff = 6240 \pm 100$~K, surface gravity $\log g = 4.16 \pm 0.20$, and iron abundance  $\fe = 0.09 \pm 0.15$.  These measured values are consistent with the KIC estimates  of $\teff = 6498 \pm 200$~K and $\log g = 4.08 \pm 0.4$ (with uncertainties estimated by \citealt{2011BL}) but are more accurate and precise because they come from high-resolution spectroscopy rather than broadband photometry. Based on the revised, cooler value for its effective temperature, KOI-1474 may indeed have a convective envelope and thus the structures in the transit light curves (Figure \ref{fig:lightcurves}) could be due to spots. Therefore spot-induced photometric variability may allow us to measure the stellar rotation period $\prot$ (\S \ref{subsec:rot}), which we can combine with other stellar parameters to infer the transiting candidate's projected spin-orbit alignment (\S \ref{subsec:align}).
\begin{figure}[htbp]
\begin{centering}
\includegraphics{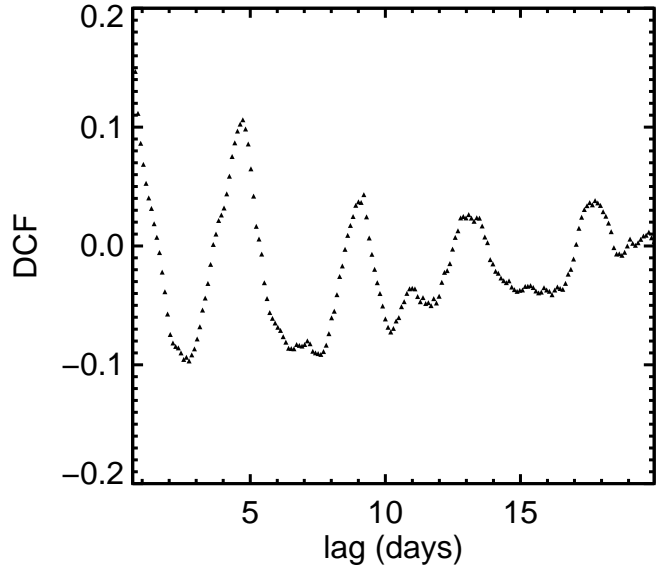}
\caption{Discrete-correlation-function (DCF, Edelson and Krolik 1988) for the long-cadence \kep Q0-Q6 KOI-1474 photometric, dataset as a function of time lag. The peak at $4.6 \pm 0.4$ days corresponds to the stellar rotation period.\label{fig:dcf}}
\end{centering}
\end{figure}
\begin{figure}[htbp]
\begin{centering}
\includegraphics{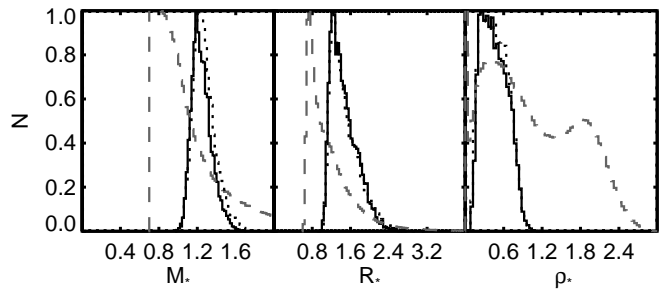}
\caption{Posteriors (solid) of stellar radius (panel 1), mass (panel 2), and density (panel 3) in solar units. The posteriors obtained from the prior alone (dashed gray) and from the data alone (dotted) are plotted in each panel, demonstrating that our data provide stronger constraints on the stellar parameters than do our priors. \label{fig:starpost}}
\end{centering}
\end{figure}

\subsection{Stellar rotation period from photometry}
\label{subsec:rot}

KOI-1474 appears to exhibit rotational photometric variability due to star spots, which cause the star to appear brighter (dimmer) as the less (more) spotted hemisphere rotates into view. We see what may be an effect of these spots in the purple, solid circle light curve marked C=377.739 in Figure \ref{fig:lightcurves}: a bump during transit consistent with a planet crossing a star spot. A periodogram (not shown) of the entire photometric dataset (Q0-Q6) exhibits a prominent cluster of peaks near 5 days. However, a periodogram is not the best tool to measure stellar rotation periods because: a) the photometric variability is non-sinusoidal, and b) the spot pattern is not expected to remain coherent over the entire 508-day dataset and thus the phase and amplitude of the best-fit sinusoid change over the data's timespan.

To obtain an optimal measurement of the stellar rotation period, we compute the discrete-correlation-function (DCF, Equation 2 of \citealt{1988E}), which was recently used to measure the rotation period of Corot-7 \citep{2009Q} and Kepler-30 \citep{2012F}. First we detrended the data with the {\tt PyKE} routine\footnote{Available at \kep Guest Observer Home: http://keplergo.arc.nasa.gov} using co-trending vectors. \citet{1999W} found that it is crucial to remove long-term trends from the time series before applying the DCF or biases may result. Then we computed the DCF using the Institut f\"{u}r Astronomie und Astrophysik T\"{u}bingen DCF routine,\footnote{Available at http://astro.uni-tuebingen.de/software/idl/aitlib/index.shtml} an IDL implementation of the DCF described in \citet{1988E}. The possible range for the DCF is -1 to 1; the amplitude is normalized such that DCF = 1 indicates perfect correlation. We plot the DCF (computed with a lag range of 0.1 days to 20 days and with 200 frequencies) as a function of time lag in Figure \ref{fig:dcf}. The DCF is highest in the region lag $<$ 0.2 days (i.e. lags that are small but greater than 0, for which the DCF =1 by definition), indicating that most of the photometric variability occurs on short timescales, most likely due to a combination of high-frequency stellar variability and instrumental noise. However, we also see lower amplitude but pronounced peaks at longer periods.

\begin{minipage}{7.0in}
\setlength{\parindent}{1cm}
\noindent 
The DCF exhibits the variations we expect due to star spots. Imagine observing the star at time t; the hemisphere in view has either more or fewer spots than the unseen hemisphere. At time $t + \prot/2$, the other hemisphere has fully rotated into view, so the flux at $t$ and $t + \prot/2$ are negatively correlated. Therefore, we interpret the negative DCF near 2 days as corresponding to half the stellar rotation period. At time $t + \prot$, we see the same hemisphere as at time t; therefore, we interpret the strong positive correlation at lag $4.6 \pm 0.4$ days as the stellar rotation period, for which uncertainty range corresponds to the width at half-maximum. The amplitude DCF = 0.1 indicates a 10$\%$ correlation between points separated in time by $\prot$. The other hemisphere rotates fully into view again at $t + 3\prot/2$, corresponding to the negative DCF at lag 7 days; at lag $2\prot$ = 9 days, the DCF is positive again. This pattern continues, and the amplitude would remain constant if the spot pattern were constant. However, the spot pattern is changing over time, so the amplitude of the correlation ``envelope'' decreases with time lag.\footnote{\begin{minipage}{7.0in}Unfortunately, the decrease in the correlation amplitude with lag implies that we are unlikely to be able to measure the stellar obliquity using the method of \citet{2011S} and \citet{2011NFF}. The spot cycle is likely shorter than the interval between subsequent transits.\end{minipage}} The measured rotation period of 4.6 days is consistent with the distribution measured for F, G, K stars by \citet{2003R}; they find that the distribution of projected rotation periods (i.e. the rotation periods measured from $\vsini$ assuming $i_s=90^\circ$) peaks at 5 days.

\subsection{Stellar density from evolution models}
\label{subsec:stellar}

The candidate's orbital eccentricity, the ultimate quantity of interest, depends weakly on the host star's density (see Paper I and references therein). Thus it is important to have an accurate, if not precise, estimate of the host star's density and, importantly, a conservative estimate of the uncertainty. For this task we use the finely-sampled YREC stellar evolution models computed by  \citet{2007T}, sampled evenly in intervals of 0.02 dex, 0.02 $\msun$, and 0.02 Gyr for metallicity $\fe$, stellar mass $\mstar$ and age $\taustar$ respectively. The model parameters are stellar age $\taustar$, mass $\mstar$, and fractional metallicity $Z$, and we wish to match the effective temperature $\teff$, surface gravity $\log g$, and $\fe$ measured spectroscopically in \S \ref{subsec:stellar}, along with their $68.2\%$ confidence ranges denoted by their ``one-sigma errors" $\{\sigma_{\teff}, \sigma_{\log g}, \sigma_{\fe}\}$, respectively. In what follows, the subscript ``spec" refers to the spectroscopically measured quantity, while quantities with no subscript are the model parameters.

Applying Bayes' theorem, the model posterior probability distribution is
\begin{eqnarray}
\label{eqn:pmodel}
\prob(\mstar, \taustar, Z | T_{\rm eff, spec},\fe_{\rm spec},\log g_{\rm spec},I) \propto \nonumber \\
\prob( T_{\rm eff, spec},\fe_{\rm spec},\log g_{\rm spec}|\mstar, \taustar, Z,I) 
\prob(\mstar, \taustar, Z | I)
\end{eqnarray}
where $I$ represents additional information available to us based on prior knowledge of the Galactic stellar population.

The first term on the right hand side (RHS) is the likelihood, which we compute by comparing the effective temperature, surface gravity, and metallicity generated by the model to the values we measured from spectroscopy:
\begin{equation}
\prob( T_{\rm eff, spec}\fe_{\rm spec},\log g_{\rm spec}| \mstar, \taustar, Z, I ) \propto \exp \left(-\frac{\chi_{\teff}^2}{2}\right) \exp\left(-\frac{\chi_{\fe}^2}{2}\right)\exp\left(-\frac{\chi_{\log g}^2}{2}\right)
\end{equation}

where
\begin{eqnarray}
\chi_{\teff}^2= \frac{\left[T(\mstar,\taustar,Z)-T_{\rm eff, spec} \right ]^2}{\sigma_{\rm Teff, spec}^2}  \nonumber \\
\chi_{\fe}^2 = \frac{\left[\fe(\mstar,\taustar,Z)-\fe_{\rm, spec} \right ]^2}{\sigma_{\fe, {\rm spec}}^2} \nonumber \\
\chi_{\log g} = \frac{\left[\log g(\mstar,\taustar,Z)-\log g_{\rm spec} \right ]^2}{\sigma_{\rm \log g, spec}^2} 
\end{eqnarray}

The second term on the RHS of Equation (\ref{eqn:pmodel}), $\prob(\mstar, \taustar, Z | I)$, is the prior information known about the model parameters. Here we make use of some additional information $I$---the galactic latitude and longitude of the Kepler field and the measured apparent Kepler magnitude of KOI-1474---to infer the relative probability of observing different types of stars. A number of factors go into this probability, including the present-day stellar mass function, the volume distribution and ages of stars along our line of sight to the Kepler field, and the Malmquist bias. Fortunately, the TRILEGAL code (TRIdimensional modeL of thE GALaxy; \citealt{2005G}) synthesizes a large body of observational, empirical, and theoretical studies to produce a model population of stars in the Kepler field that are consistent with KOI-1474's apparent Kepler magnitude $K_P = 13.005 \pm 0.030$ \citep{2010B} and Galactic coordinates. From this model population, we use a Gaussian kernel density estimator to compute a three-dimensional density function for the prior $\prob(\mstar, \taustar, Z | I)$.

Each combination of \citet{2007T} model parameters --- $(\mstar, \taustar, Z)$ --- has a corresponding $\rstar$ and $L_\star$, and we calculate the corresponding stellar density $\rhostar~=~\frac{\mstar}{M_\odot}~(\frac{R_\odot}{\rstar})^3~\rho_\odot$. We compute the star's absolute Kepler bandpass magnitude $K_{P, {\rm absolute}}$ through the follow steps: we transform $L_\star$ into a $V$ magnitude using a bolometric correction, transform $V$ into the absolute Sloan magnitude $g$, and compute the distance modulus using the difference \end{minipage}
\clearpage \noindent between the absolute $g$ magnitude and the apparent $g$ magnitude from the KIC \citep{2010B}. Then we apply the distance modulus to the apparent $K_P$ to obtain $K_{P, {\rm absolute}}$. Thus we can transform the model posterior $\prob(\mstar, \taustar, Z | T_{\rm eff, spec},\fe_{\rm spec},\log g_{\rm spec},I) $ into posteriors for the stellar stellar properties $\mstar$, $\taustar$, $\rstar$, $\rhostar$, $L_\star$, and $K_{P, {\rm absolute}}$ (Table \ref{tab:starparams}, column 3). In Figure \ref{fig:starpost} we plot the resulting posteriors for  $\mstar$, $\rstar$, and $\rhostar$. We also plot the same distributions obtained from the data alone and from the priors\footnote{The $\mstar$  prior probability appears truncated below $\mstar=0.78$ in Figure \ref{fig:starpost} because we only compute \citet{2007T} models above this value. However, the likelihood completely rules out stars with $\mstar < 1 M_\odot$. } alone; evidently most of the constraint comes from the data (i.e. the spectroscopic quantities).

\begin{deluxetable*}{rrr}
\tabletypesize{\footnotesize}%
\tablecaption{Stellar Parameters for KOI 1474 \label{tab:starparams}}
\tablewidth{0pt}
\tablehead{
\colhead{Parameter}    & \colhead{Value\tablenotemark{a}}}
\startdata
&Measured&Derived from model\\
\hline
\\
Right ascension, RA (hour,J2000) 						& 19.694530							\\
Declination, Dec (degree,J2000) 						& 51.184800							\\
Projected rotation speed, $v_{\rm rot} \sin i_s$ [km s$^{-1}$] 		& 13.6$ \pm$0.5						\\
Stellar effective temperature, $\teff$ [K] 			& 6240$\pm$100 		&6230$\pm$100	 \\
Iron abundance, $\mbox{[Fe/H]}$ 					&0.09 $\pm$0.15		&0.00 $^{+0.16}_{-0.12}$	\\
Surface gravity, $\log(g [$cms$^{-2}$] 				&4.16$\pm$0.20		&4.23$^{+0.13}_{-0.16}$	\\
Limb darkening coefficient, $\mu_{1}\tablenotemark{b}$ 	&					& 0.320 $\pm$ 0.015 	\\
Limb darkening coefficient, $\mu_{2}\tablenotemark{b}$ 	&					& 0.304 $\pm$ 0.007 	\\
Main sequence age, $\taustar$ [Gyr] \tablenotemark{c} 					& 					&2.8$^{+1.3}_{-1.2}$			\\
Stellar mass, $M_{\star}$ [$M_{\odot}$]	 \tablenotemark{c}	&					&  1.22$^{+0.12}_{-0.08}$ 		\\
Stellar radius, $R_{\star}$ [$R_{\odot}$] 		&					&  1.40$^{+0.37}_{-0.21}$ 		\\
Stellar density, $\rho_{\star}$ [$\rho_{\odot}$]	&					& 0.44$^{+0.26}_{-0.20}$					\\
Stellar luminosity, $L_{\star}$ [$L_{\odot}$] 		&					&  2.7$^{+1.6}_{-0.8}$ 					\\
Apparent Kepler-band magnitude, $K_P$	& 13.005 $\pm$ 0.030\\
Absolute Kepler-band magnitude, $K_{P, {\rm absolute}}$				&					& 3.6$^{+0.4}_{-0.5}$ 					\\
Distance (kpc)							&						& 0.78$^{+0.23}_{-0.13}$  \\
Rotation period, $\prot$ [days] 				&4.6 $\pm$ 0.4& 		\\
Rotation speed, $v_{\rm rot}$[km s$^{-1}$] 	&					& 14.7$^{+2.6}_{-1.0}$ 		\\
Sine of stellar spin axis inclination angle, $\sin i_s$				&					&0.93$^{+0.06}_{-0.14}$ \\
Stellar spin axis inclination angle, $i_s$[degree]				&					&69$^{+14}_{-17}$  \\ 
Deviation of stellar spin axis from edge-on, $|90-i_s$ [degree]		&					&21$^{+17}_{-14}$ \\ \\
\enddata
\tablenotetext{a}{The uncertainties represent the 68.3\% confidence interval of the posterior distribution.}
\tablenotetext{b} {Sing 2010}
\tablenotetext{c} {A prior was imposed on this parameter.}
\end{deluxetable*}

\begin{deluxetable*}{rrr}
\tabletypesize{\footnotesize}%
\tablecaption{Planet Parameters for KOI 1474.01 \label{tab:planparams}}
\tablewidth{0pt}
\tablehead{
\colhead{Parameter}    & \colhead{Value\tablenotemark{a}}}
\startdata
& Circular fit  															& Eccentric fit \\
\hline
\\
Average orbital period, $P$ [days]\tablenotemark{b}   		&  69.7339$\pm$0.0016 					& 69.7340$\pm$0.0015 				\\
Average mid transit epoch, $T_{c}$ [days] [BJD-2455000]   	& 238.273$\pm$0.011					& 238.273$\pm$0.010				\\
Mid transit epoch of transit 1, $T_1$ [days] [BJD-2455000]     	& -40.6701$\pm$0.0008  					& -40.6702$\pm$0.0009  \\
$T_2$ [days] [BJD-2455000]  	&  29.0600$\pm$0.0006  					&  29.0600$\pm$0.0007 \\
$T_3$ [days] [BJD-2455000]  	&  98.7647$\pm$0.0006  					&  98.7647$\pm$0.0007 \\
$T_4$ [days] [BJD-2455000]   	&168.5752$\pm$0.0006  					&168.5752$\pm$0.0007 \\
$T_5$ [days] [BJD-2455000]  	&238.3146$\pm$0.0005  					&238.3146$\pm$0.0007 \\
 $T_6$ [days] [BJD-2455000]  	&308.0092$\pm$0.0008  					&308.0092$\pm$0.0009 \\
$T_7$ [days] [BJD-2455000]  	&377.7250$\pm$0.0006  					& 377.7250$\pm$0.0007 \\
$T_8$ [days] [BJD-2455000] 	&447.4555$\pm$0.0006  					&447.4555$\pm$0.0007 \\
Planet-to-star radius ratio, $R_{p}/R_{\star}$   				&0.0618 $^{+0.0007}_{-0.0003}$ 		& 0.0617 $^{+0.0006}_{-0.0004}$ 		\\
Stellar density, $\rho_{\star}$    							&9.2 $^{+0.4}_{-1.6}$ 				 	& 0.36\tablenotemark{c}$^{+0.30}_{-0.10}$	\\
Orbital inclination, $i$ [degree]  						&89.93 $^{+0.05}_{-0.08}$		 		& 89.2 $^{+0.4}_{-1.3}$				\\
Limb darkening coefficient, $\mu_{1}\tablenotemark{c} $ 			& 0.314$^{+0.018}_{-0.012}$				& 0.311$^{+0.016}_{-0.012}$			\\
Limb darkening coefficient, $\mu_{2}\tablenotemark{c} $ 			& 0.302 $^{+0.006}_{-0.008}$				& 0.304$^{+0.005}_{-0.009}$				\\
Impact parameter, $b$ 								& 0.18$^{+0.21}_{-0.12}$		 			& 0.14$^{+0.25}_{-0.09}$				\\
Planetary radius, $R_{p}$ [$R_{\oplus}$] 				&								 	&9.5$^{+2.4}_{-1.4}$				\\
Normalized red noise, $\sigma_r$						& 0.00005$^{+0.00007}_{-0.00003}$  		& 0.00007$^{+0.00005}_{-0.00005}$ 	\\
Normalized white noise, $\sigma_w$						& 0.000131$^{+0.000010}_{-0.000004}$  	& 0.000134$^{+0.000007}_{-0.000007}$ 	\\
Eccentricity, $e$ 						&									&0.81$^{+0.10}_{-0.07}$				\\
Orbital period after tidal circularization, $P_{\rm final}$						&									&14$^{+6}_{-10}$					\\
Line-of-sight spin-orbit angle, $|i-i_s|$	[degree]						&	&$21.^{+17}_{-14}$ \\
\enddata
\tablenotetext{a}{The uncertainties represent the 68.3\% confidence interval of the posterior distribution.}
\tablenotetext{b}{$P$ and $T_{c}$ are determined from a linear fit to the transit times. The uncertainty in $T_{c}$ is the median absolute deviation of the transit times from this ephemeris; the uncertainty $P$ is this quantity divided by the number of orbits between the first and last observed transits.}
\tablenotetext{c} {A prior was imposed on this parameter.}
\end{deluxetable*}

The derived density for KOI-1474, 0.44$^{+0.26}_{-0.20} \rho_\odot$, has an uncertainty range encompassing the KIC value of 0.26 $\rho_\odot$ \citep{2010B}. The star is significantly less dense than the value of $6 \rho_\odot$ derived from $a/\rstar$ in the table of candidates \citep{2011BKB,2012B} . Therefore, planet candidate KOI-1474.01 is likely to have a large eccentricity, which we will measure in \S \ref{sec:ecc}. Fortunately, as shown as Paper I, even the loose constraint on the stellar density derived here will result in a precise measurement of the candidate's large orbital eccentricity.
\begin{figure}[htbp]
\begin{centering}
\includegraphics[width=3.5in]{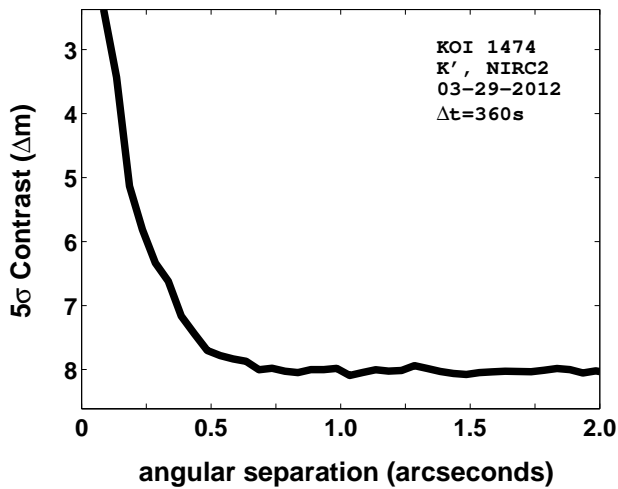}
\caption{Sensitivity to off-axis sources in the immediate vicinity of KOI-1474 using adaptive optics imaging observations with NIRC2 at Keck in the K'-band ($\lambda_c=2.12 \mu m$). \label{fig:ao}}
\vspace{0.3 in}
\end{centering}
\end{figure}
\section{False Positive Probability}
\label{sec:fpp}

\begin{figure}
\begin{centering}
\includegraphics[width=3.5in]{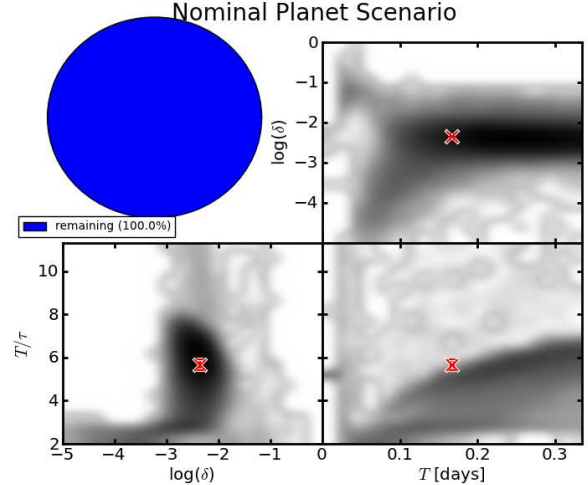} 
\caption{Three-dimensional probability distribution for the trapezoidal shape parameters (depth $\delta$, duration $T$, and ``slope'' $T/\tau$) for the nominal planet scenario. The distributions are generated by simulating a statistically representative population (see \citealt{2012M}, \S 3.1) for the scenario and fitting the shape parameters to each simulated instance. Each population begins with 100,000 simulated instances, and only instances that pass all available observational constraints are included in these distributions. In this case, no additional observational constraints are available so the $100\%$ of the distribution remains. The transit's shape parameters $\delta$, $T$, and $T/\tau$ are marked on each plot with an ``X'' denoting the the median of an MCMC fit. \label{fig:fppplan}}
\end{centering}
\end{figure}

\begin{figure}
\begin{centering}
\includegraphics[width=3.5in]{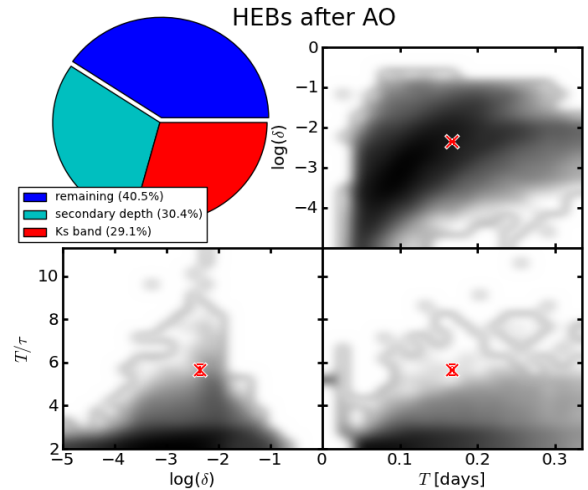}
\caption{Same as Figure \ref{fig:fppplan} for the HEB scenario. In this case, the upper-limit of 200 ppm we place on the secondary eclipse depth eliminates $30.4\%$ of the distribution and limits from the Ks-band adaptive optics image eliminate $29.1\%$ of the distribution, leaving 40.5$\%$ remaining. \label{fig:fppheb}}
\end{centering}
\end{figure}

\begin{figure}
\begin{centering}
\includegraphics[width=3.5in]{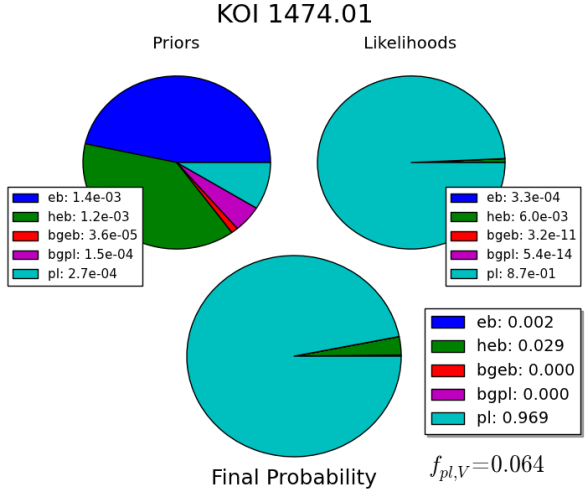}
\caption{Prior (top left), likelihood (top right), and final (bottom) probabilities for four false positive scenarios --- an undiluted eclipsing binary (``eb"), hierarchical eclipsing binary (``heb"), background eclipsing binary (``bgeb"), and background planet (``bgpl"). The priors and likelihoods are computed following \citet{2012M}. Each final probability is the product of the scenario's prior and likelihood, normalized so that the total probabilities sum to 1. The quantity $f_{\rm pl,V}$ indicates the specific occurrence rate for planets of this size that we would need to assume in order for the FPP to be less than $0.5\%$. Since this rate, $f_{\rm pl,V} = 6.4\%$, is higher than our assumed $f_{\rm pl} = 0.01$,  we do not consider the candidate validated. \label{fig:fppsum}}
\end{centering}
\end{figure}

Although a transiting planet may cause the photometric signal observed in light curves (Figure \ref{fig:lightcurves}), any of several scenarios involving stellar eclipsing binaries might cause a similar signal.  This is the well-known problem of astrophysical false positives for transit surveys \citep[e.g.][]{2003B,2011TF}.  Traditionally transiting planets have been confirmed through detection of their radial velocity (RV) signals. However, the {\it Kepler} mission has necessitated a different paradigm: one of {\it probabilistic validation}.  If the false positive probability (FPP) of a given transit signal can be shown to be sufficiently low (e.g.~$\ll 1\%$), then the planet can be considered {\it validated}, even if not dynamically confirmed. Here we attempt to validate KOI-1474.01 but find a 3.1$\%$ probability that the signal is due to an astrophysical false positive.

At first glance, the short duration of KOI-1474.01's transit (Section \ref{sec:lightcurve}) causes particular concern: the signal could be a transit or eclipse of an object orbiting a smaller, blended star, which would make the duration more in line with that expected for a circular orbit. In order to calculate the FPP for KOI-1474.01, we follow the procedure outlined in \citet{2012M}, which incorporates simulations of realistic populations of false positive scenarios, the KIC colors, the measured spectroscopic stellar properties, and a descriptive, trapezoidal fit to the photometric data.  

To place constraints on blending by searching for nearby sources, we obtained adaptive optics images of KOI-1474 on March 29, 2012 using NIRC2 (PI: Keith Matthews) at the 10m Keck II telescopes. KOI-1474 is sufficiently bright to serve as its own natural guide star ($K_P = 13.005$) and therefore does not require the use of a laser to correct for wavefront errors introduced by the Earth's atmosphere. Our observations consist of 18 dithered images (10 coadds per frame, 2 seconds per coadd) taken in the $K^\prime$ filter ($\lambda_c=2.12 \mu$m). We used NIRC2's narrow camera mode, which has a platescale of 10 mas / pix, to provide fine spatial sampling of the stellar point-spread function. 

Raw frames were processed by cleaning hot pixels, flat-fielding, subtracting background noise from the sky and instrument optics, and coadding the results. No off-axis sources were noticed in individual frames or the final processed image. Figure \ref{fig:ao} shows the contrast levels achieved from our observations. Our diffraction-limited images rule out the presence of contaminants down to $\Delta K'=5$~mag and $\Delta K'=8$~mag fainter than the primary star for separations beyond 0.2'' and 0.7'' respectively.  

\begin{figure}
\begin{centering}
\includegraphics{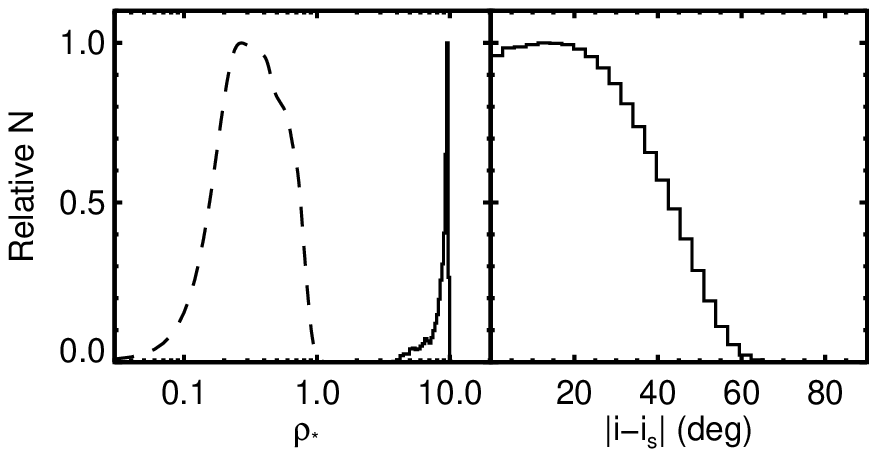}\\
\includegraphics{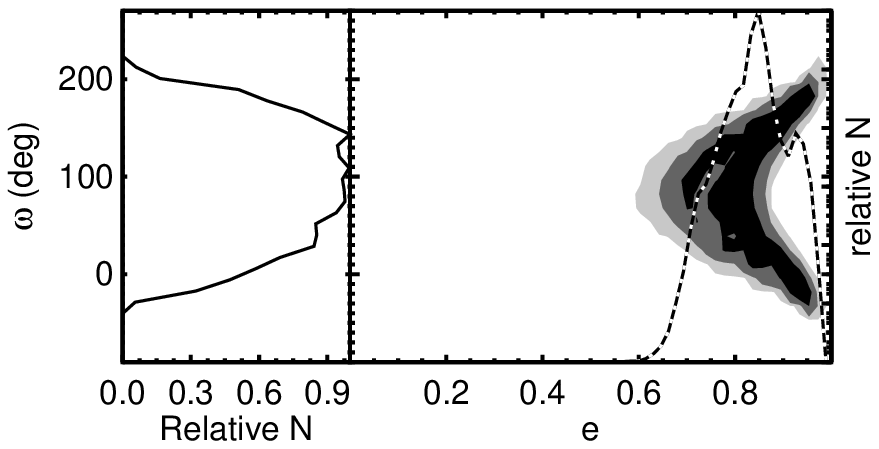}
\caption{Top left: $\rho_{\rm circ}$ obtained from circular fit to the transit light curve (solid) and posterior for $\rhostar$ from \S \ref{subsec:stellar} (dashed); since the host star is not highly dense (i.e. the two posteriors do not overlap), the planet's orbit must be highly eccentric. Top right: Posterior for projected spin-orbit alignment from an eccentric fit to transit light curve, imposing a prior on $\rhostar$. Bottom left: Posterior distribution $\omega$ from an eccentric fit to transit light curve, imposing a prior on $\rhostar$. Bottom right: Joint posterior for $\omega$ vs. $e$. The black (gray, light gray) contours represent the $\{68.3,95,99\}$\% probability density levels (i.e. 68$\%$ of the posterior is contained within the black contour) Over-plotted as a black-and-white dotted line is a histogram of the eccentricity posterior probability distribution marginalized over $\omega$.  \label{fig:eccpost} }
\end{centering}
\end{figure}

We plot the probability distributions for the nominal planet scenario in Figure \ref{fig:fppplan}, as well as for the most likely alternative to a transiting planet: an hierarchical eclipsing binary (HEB) (Figure \ref{fig:fppheb}), in which KOI-1474 has a wide binary companion of comparable brightness (within a few magnitudes) that is being eclipsed by a small tertiary companion. The probability of the HEB scenario is 2.3\%. In Figure \ref{fig:fppsum}, we\begin{minipage}{7.0in}
\setlength{\parindent}{1cm}
\noindent  summarize the prior, likelihood, and total probability of the nominal transiting planet scenario compared to that of several false positive scenarios. The FPP is:

\begin{equation}
{\rm FPP} = \frac{L_{FP} } {L_{FP} + \frac{f_P}{0.01} L_{TP}} = \frac{(0.002+0.029+0.000+0.000)}{(0.002+0.029+0.000+0.000)+\frac{0.01}{0.01} 0.969} = 0.031
\end{equation}
\noindent where $L_{FP}$ is the sum of the probabilities of the false-positive scenarios, $L_{TP}$ is the probability of the nominal planet scenario, and $f_P$ is the assumed specific occurrence rate\footnote{\begin{minipage}{7.0in}The assumed 1\% occurrence rate is motivated by the debiased 1\% occurrence rate for hot Jupiters in the RV sample \citep{2012W}. In order to produce a FPP of than 0.5\%, $f_p$ would have be greater than 6.4\%. See \citet{2012M} for a discussion of specific planet occurrence rates.\end{minipage}} for planets between 5.7 and 11.3 $R_\oplus$. Although this FPP is low, we do not consider it sufficiently low to validate the planet. In the analysis following in the remainder of the paper, we assume that KOI-1474.01 is a planet and refer to it as ``planet," but in fact it remains a candidate planet.  We are conducting a radial-velocity follow-up campaign of this target to confirm this candidate by measuring its mass.

\section{The highly eccentric orbit of KOI-1474.01}
\label{sec:ecc}

In \S \ref{sec:star}, we revised the stellar properties of KOI-1474 and found that the star's density indicates that the (validated) planet's orbit is highly eccentric. To quantify the eccentricity, we now model the light curves (Figure \ref{fig:lightcurves}) with the Transit Analysis Package software \citep[TAP,][]{2011G} to obtain the posterior distribution for the eccentricity and other transit parameters (\S \ref{subsec:ecc}), using the technique described in Paper I. In \S \ref{subsec:align}, we place constraints on the spin-orbit alignment based on stellar properties measured in \S \ref{subsec:stellar}. In \S \ref{subsec:ttvs}, we assess the observed TTVs and explore the nature of the third-body perturber.

\subsection{Fitting orbital parameters to the light curve}
\label{subsec:ecc}

Here we measure KOI-1474.01's orbital parameters, including eccentricity, from the transit light curves (Figure \ref{fig:lightcurves}). We use TAP to fit a \citet{2002MA} light curve model, employing the wavelet likelihood function of \citet{2009C}. We replace the parameter $a/\rstar$ with $\rhostar$ \citep[][Equation 30]{2010W} in the limit that $(\mstar+M_p)/(\frac{4}{3} \pi \rstar)^3 \rightarrow \rhostar$, but transform $\rhostar$ into $a/\rstar$ to compute the light curve model. Using the spectroscopic stellar parameters measured in \S \ref{subsec:spec} (Table \ref{tab:starparams}, column 2), we calculate the limb darkening coefficients $\mu_1$ and $\mu_2$ and their uncertainties with the table and interpolation routine provided by \citet{2010S}. In all the orbital fits  discussed 
\end{minipage}
\clearpage
\noindent 
\begin{minipage}{7.0in}
\setlength{\parindent}{1cm}
\noindent here, we impose normal priors on $\mu_1$ and $\mu_2$ (Table \ref{tab:starparams}), which are well-measured for the Kepler bandpass. We also verified that uniform priors on the limb darkening yield consistent results (with slightly larger uncertainties) for all the orbital fits we perform. The other light curve parameters we fit for are the mid-transit time of each light curve $T$, the planet-to-stellar radius ratio $R_p/\rstar$, the fractional white noise $\sigma_w$, the red noise $\sigma_r$, the inclination $i$, and the argument of periapse $\omega$, with uniform priors on each of these quantities. 

Finally, to speed up the fit convergence, we explore parameter space using the parameter $g$ instead of the planet's orbital eccentricity $e$. The parameter $g$ corresponds approximately to the ratio of the observed transit speed to the speed expected of a planet with the same period but $e=0$:

\begin{equation}
\label{eqn:g}
g(e,\omega)=\frac{1+e\sin\omega}{\sqrt{1-e^2}} = \left(\frac{\rho_\star}{\rho_{\rm circ}}\right)^{1/3}
\end{equation}

We impose a prior on g to maintain a uniform eccentricity prior (see \S 3.3.1 of Paper I for further details):

\begin{eqnarray}
\label{eqn:gprior}
\prob (g ) =  \frac{\sin^2\omega\left(\sin^2\omega-1\right)+g^2\left(1+\sin^2\omega\right)\pm2g\sin\omega\sqrt{\sin^2\omega-1+g^2}}{\sqrt{\sin^2\omega-1+g^2}\left(g^2+\sin^2\omega\right)^2}
\end{eqnarray}
\noindent for which the $+$ corresponds to $g >1 $ and the $-$ to $g<1$. We transform $g$ into $e$ to compute the light curve model.  

First we fit a circular orbit (Table \ref{tab:planparams}, column 2), fixing $e=0$ and leaving free $\rhostar$, to which we refer as  $\rho_{\rm circ}$. We find that: 1) although we only have long-cadence data for KOI-1474 (Figure \ref{fig:lightcurves}), $\rho_{\rm circ}$ and the impact parameter $b$ are separately well-constrained (see also \S 4.2 of Paper I for a discussion of long-cadence data), and 2)the $\rhostar$ posterior computed from stellar properties in \S \ref{subsec:stellar} ($\rhostar = 0.44^{+0.26}_{-0.20} \rho_{\odot}$) falls far outside the transit light curve posterior distribution for $\rho_{\rm circ}$ ($\rhostar = 9.2 ^{+0.4}_{-1.6} \rho_{\odot}$, Figure \ref{fig:eccpost}, top left panel), where the uncertainties indicate the 68.3\% confidence interval.  Thus a circular fit is inconsistent with our prior knowledge of the stellar parameters.  

Because the eccentricity depends only weakly on the assumed stellar density, the eccentricity measurement we are about to perform is relatively robust to errors in the assumed stellar density. When $\rhostar > \rho_{\rm circ}$, the transiting planet has a minimum eccentricity obtained by setting $\omega = \pi/2$ in Equation (\ref{eqn:g}) (i.e. the planet transits at periapse). Imagine that $\rhostar$ were biased or in error. The fractional change in $e_{\rm min}$ would be:
\begin{equation}
\label{eqn:de}
\frac{\Delta e_{\rm min}}{e_{\rm min}} = \frac{4}{3\left[\left(\frac{\rho_\star}{\rho_{\rm circ}}\right)^{2/3}-\left(\frac{\rho_\star}{\rho_{\rm circ}}\right)^{-2/3}\right]} \frac{\Delta (\frac{\rho_\star}{\rho_{\rm circ}})}{(\frac{\rho_\star}{\rho_{\rm circ}})}
\end{equation}
The ratio $\frac{\rho_\star}{\rho_{\rm circ}} = \frac{9.2}{0.44} = 21$, corresponding to $e_{\rm min} = 0.77$ and $\frac{\Delta e_{\rm min}}{e_{\rm min}}  = 0.18 \frac{\Delta (\frac{\rho_\star}{\rho_{\rm circ}})}{(\frac{\rho_\star}{\rho_{\rm circ}})}$. So if the stellar density were biased upward by $10\%$, the minimum eccentricity would be biased upward by only 1.8$\%$. See \S 3.1 and \S 4.1 Paper I for a detailed exploration of how the stellar density's assumed probability distribution affects the eccentricity measurement.

Next we fit the light curve allowing the planet to have an eccentric orbit (Table \ref{tab:planparams}, column 3) and using the stellar density posterior from \S \ref{subsec:stellar} as the stellar density prior for the light curve fit. As argued in Paper I (Section 3), an MCMC exploration --- as implemented in TAP --- naturally accounts for the transit probability and marginalizes over the uncertainties in other parameters. Even though $e$ and $\omega$ are degenerate for a given $g$ (Equation~\ref{eqn:g}), there is a lower limit on e, and the posterior falls off gradually, as $e\rightarrow 1$ and the range of possible $\omega$ satisfying Equation \ref{eqn:g} narrows. The posterior distributions for $e$ and $\omega$ are plotted in Figure \ref{fig:eccpost}. We measure $e = 0.81^{+0.10}_{-0.07}$. For comparison, if we had set the stellar density prior to be uniform between $0.1 \rho_\odot - 0.2 \rho_\odot$ ($0.6 \rho_\odot - 1.2 \rho_\odot$), we would measure $e = 0.90^{+0.03}_{-0.03}$ ($e = 0.73^{+0.15}_{-0.09}$).

By conservation  of angular momentum, this planet would attain a final period $\pfinal (1-e^2)^{3/2}  =  14^{+9}_{-10}$ days if it were to undergo full tidal circularization. In \S \ref{sec:protofail}, we will discuss whether the planet is best classified as a proto-hot Jupiter --- likely to circularize over the star's lifetime and achieve a short-period orbit --- or a failed-hot Jupiter, just outside the reach of fast tidal circularization.

\subsection{Constraints on spin-orbit alignment}
\label{subsec:align}

Whatever process perturbed KOI-1474.01 onto an eccentric orbit may have also tilted the planet's orbit from the plane in which it formed. With a temperature of $6240 \pm 100$~K (Section \ref{subsec:spec}), KOI-1474 sits right on the 6250 K boundary between hot stars with high obliquities and cool stars with well-aligned planets \citep{2010WF}. However, if 1) cool stars have low obliquities because their hot Jupiters have realigned the star's outer convective layer, as proposed by \citet{2010WF}, and 2) KOI-1474.01 is a failed-hot Jupiter, with a tidal dissipation rate too low to experience significant circularization over KOI-1474's lifetime, then KOI-1474.01 may have also not yet realigned KOI-1474's outer layer.
\end{minipage}
\clearpage
\noindent 
\begin{minipage}{7.0in}
\setlength{\parindent}{1cm}
\noindent 
Ultimately we will wish to determine $\psi$, the total misalignment between the orbit normal and the host star spin axis, from three measured projected angles \citep{2009F,2010Schlaufman}: $i$, the inclination between the planet's orbit and the observer's line of sight; the sky-projected spin-orbit angle $\lambda$; and $i_s$, the inclination between the stellar spin axis and the line of sight. We measured $i$ from the transit light curve in \S \ref{subsec:ecc} (Table \ref{tab:planparams}). The sky-projected spin-orbit angle $\lambda$ could one day be measured via the Rossiter-McLaughlin (RM) effect \citep{1924M,1924R,2000Q},  the change in the observed radial velocity as a transiting planet blocks portions of the star rotating toward or away from the observer. The effect has a maximum amplitude of about 50~m/s  (\citealt{2010W},  Equation 40) However, because KOI-1474.01's transits can occur early or late by over an hour, RM measurements of KOI-1474.01 will remain challenging until the TTV pattern ``turns over'' in future Kepler observations, allowing us to predict future transits to much higher precision (Section \ref{subsec:ttvs}). We can measure the third projected angle, $i_s$, from $(\vsini)_{\rm spec}$ (Section \ref{subsec:spec}) and the posteriors of $\prot$ (Section \ref{subsec:rot}) and $\rstar$ (Section \ref{subsec:stellar}), an approach that was recently applied by \citet{2012HS} to fifteen KOI systems. KOI-1474's rotational velocity is $v_{\rm rot} = \frac{2\pi \rstar}{P_{\rm rot}}$ and we have measured the projected rotational velocity $(\vsini)_{\rm spec}$. Therefore we can find the angle of that projection, $i_s$. According to Bayes theorem:
\begin{equation}
\label{eqn:rotmodel_post}
\prob\left(\prot, \rstar,  i_s  | (\vsini)_{\rm spec} \right) = \prob\left((\vsini)_{\rm spec} | \prot, \rstar,   i_s \right) \prob\left(\prot, \rstar, i_s\right).
\end{equation}
\noindent The prior, $\prob\left(\prot, \rstar,  i_s\right)$, is the product $$\prob\left(\prot, \rstar,  i_s\right) = \prob(\prot) \prob( \rstar) \prob ( i_s)$$ where $\prob(\prot)$ is a normal distribution with mean 4.6 m/s and standard deviation 0.4 m/s (\S \ref{subsec:rot}) and $ \prob( \rstar) $ is the posterior from \S \ref{subsec:stellar}. Assuming stellar spin axes are randomly oriented throughout the Galaxy, the distribution of $ \cos i_s$ is uniform and thus $\prob ( i_s) = \frac{1}{2} \sin i_s$. 

Next, we integrate Equation \ref{eqn:rotmodel_post} over $\prot$ and $ \rstar$ to  obtain the stellar inclination $i_s$ conditioned on our measured projected rotational velocity $\vsini$. 
\begin{equation}
\label{eqn:is_post}
\prob\left(i_s  | (\vsini)_{\rm spec} \right) = \int \int \prob\left((\vsini)_{\rm spec} | \prot, \rstar,   i_s \right) \prob\left(\prot, \rstar, i_s\right) d\prot d\rstar
\end{equation}
\noindent As a practical implementation of Equation (\ref{eqn:is_post}) we randomly draw $\prot$ and $\rstar$ from the distributions calculated in \S \ref{subsec:rot} and \S \ref{subsec:stellar} respectively and $ i_s$ from a uniform distribution of $\cos i$ between 0 and 1. Drawing from these respective distributions is equivalent to creating a grid in these parameters and subsequently downsampling according to the prior probabilitities. Then we compute the likelihood 
\begin{equation} 
\label{eqn:probv}
\prob\left((\vsini)_{\rm spec} | \prot, \rstar,   i_s \right) = \exp\left[-\left(\frac{2\pi \rstar}{P_{\rm rot}} \sin i_s - (\vsini)_{\rm spec}\right)^2/\left(2 \sigma_{(\vsini)_{\rm spec}}^2\right)\right]
\end{equation}
\noindent where $(\vsini)_{\rm spec}$ =13.6 m/s and $\sigma_{(\vsini)_{\rm spec}}$ = 0.5 m/s (Section \ref{subsec:spec}). Then we select a uniform random number between 0 and 1; if the uniform random number is less than $\prob\left((\vsini)_{\rm spec} | \prot, \rstar,   i_s \right) $ (Equation \ref{eqn:probv}), we include the model $\left( \prot, \rstar,   i_s \right)$ in the posterior. We repeat drawing $\left( \prot, \rstar,   i_s \right)$ models until we have thousands of models that comprise the posterior.

We measure a projected angle for the stellar spin axis $i_s = 69^{+14}_{-17}$ degrees. Combining the posterior of $i_s$ with the posterior of the planet's inclination $i$ (Section \ref{subsec:ecc}),we obtain $|i - i_s|= 21^{+17}_{-14}$, for which the total uncertainty is dominated by the uncertainty in the stellar radius. We list these angles in Table \ref{tab:planparams}, and plot the posterior for the line-of-sight spin-orbit angle $|i - i_s|$ in Figure \ref{fig:eccpost} (top right panel). Our posterior distribution is consistent (within 2 $\sigma$) with close alignment, yet allows misaligned configurations as well. We also caution that differential rotation may cause systematic errors in the measured alignment, depending on the latitude of the spots (see \citealt{2012HS}, Section 5.3 for a detailed discussion). Furthermore, the line-of-sight spin-orbit angle $|i - i_s|$ offers no constraint on whether the planet's orbit is prograde or retrograde. However, two types of future follow-up observations will allow us to better constrain the planet's orbit in three dimensions. First, additional constraints on the planet's orbit through radial-velocity measurements will in turn constrain the stellar radius, providing a more precise measurement of $|i - i_s|$. To this end we are currently conducting a Doppler follow-up program at Keck with HIRES. Second, from the measurement of the sky-projected spin-orbit angle $\lambda$ via the RM effect, the total spin-orbit angle $\psi$ can be computed by combining $\lambda$ with a refined line-of-sight measurement $|i-i_s|$.
\subsection{Transit Timing Variations}
\label{subsec:ttvs}

The light curves in Figure \ref{fig:lightcurves} reveal large variations in the mid-transit times of KOI-1474.01, which may be caused by perturbations from another planet or sub-stellar companion. If KOI-1474.01 underwent HEM, this perturber may have been responsible. Table \ref{tab:planparams} displays the mid-transit times from the orbital fits performed in Section \ref{subsec:ecc}. There the best-fitting linear ephemeris is also given, from which the times deviate significantly. In Figure~\ref{fig:oc}, we plot an observed minus calculated (O-C) diagram of the observed transit time minus the transit time calculated from a constant orbital period.
The scale and sharpness of the features in Figure \ref{fig:oc} suggest a nearby giant planet or brown dwarf perturber. We assume this perturber is on an exterior orbit, as KOI-1474.01's eccentric orbit leaves little dynamical room interior to itself.  We are undertaking a radial-velocity follow-up campaign (Johnson et al. 2013, in prep) that may allow us to rule out an interior, Jupiter-mass companion.

\end{minipage}
\clearpage

The ``jump'' in the O-C diagram likely corresponds to the periapse passage of an eccentric companion \citep{2003BE,2004B,2005A,2011BC}. Throughout its orbit, this perturbing companion creates a tidal force on the orbit of the transiting planet. If the companion's orbit is exterior to and within the plane of the transiting planet's, the tidal force increases the inner planet's orbital period or, equivalently, decreases the effective mass of the central star (see Section 4 of \citealt{2005A} for a detailed derivation). The tidal force varies with the distance between the perturber and star and is strongest when the perturber is at periapse. Therefore, as the perturber approaches periapse, the transiting planet's orbital period lengthens, causing later and later transit arrival times, corresponding to the discontinuity seen in Figure \ref{fig:oc}. The period of the TTV cycle corresponds to the perturbing planet's orbital period. The amplitude is set by the change in the tidal force (a combination of the perturbing planet's mass and periapse distance, which is a function of the eccentricity and orbital period). The sharpness of the O-C depends on the perturber's eccentricity --- whether the perturbation is the flyby of a companion on a highly eccentric orbit or the gradual approach of a moderately eccentric companion. The transiting planet's orbital eccentricity also subtly affects the shape of the O-C diagram, as explored in detail by \cite{2011BC}. Our Figure~\ref{fig:oc} has a similar appearance to the TTVs produced by \cite{2011BC}'s analytical and numerical models of eccentric, hierarchal systems.

\begin{figure*}[htbp]
\begin{centering}
\includegraphics{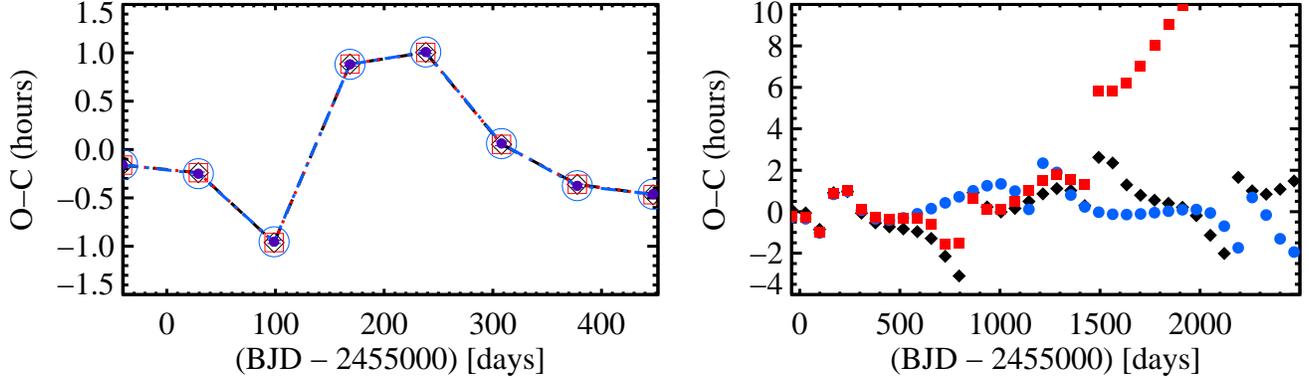}
\caption{Left: Observed mid transit times (purple dots) of the eight transits of 1474.01 with subtracted best fit linear ephemeris from the \S \ref{subsec:ecc} transit light curve model (Table \ref{tab:planparams}, column 3). TTV predictions from the first (solid black, open diamonds), second (red, open squares), and third (blue, open circles) dynamical model in Table \ref{tab:ttvfits}. All three models match the data well. Right: Same models as left plotted over longer timespan; the models differ in their predictions for future O-C variations. \label{fig:oc}}
\end{centering}
\end{figure*}

Currently we do not have a long enough TTV baseline to uniquely model the perturbing companion, as \citet{2012N} achieved for the system KOI-857.  Since ``jumps" in the O-C diagram correspond to the perturber's periapse passage and we have only seen one such jump, apparently the current TTVs cover less than one orbit of the outer companion. Therefore we cannot well constrain the outer body's orbital period. The TTV amplitude -- set by the tidal force on the transiter's orbit -- is well constrained but depends on the perturber's mass, orbital period, and eccentricity; therefore we expect to find degeneracy among these quantities. Furthermore, the tidal force on the transiter's orbit depends on the mutual inclinations of the bodies. The tide due to a polar position for the perturber would \emph{decrease} the transiter's orbital period; averaging over the bodies' positions, a very inclined perturber could be more massive and yet produce a comparable amplitude perturbation.

Here we explore a subset of all possible parameters for the perturbing planet. With only eight transit times (Table \ref{tab:planparams}), we have a great amount of freedom in the fits, but it is still of interest whether or not a physical model of a perturber can fit these data.\footnote{For example, \citet{2012N} demonstrated that, as expected, they could not find a physically plausible model when they scrambled their TTVs.  Failure to find an orbital model that reproduces the observed TTVs would cast suspicion on our interpretation that they are the signature of an unseen companion.} Thus we proceed with direct 3-body fits to the data. We do not expect the TTVs to be sensitive to the mass of the transiting planet or the host star \citep{2011BC,2012N} so we fix  $M_{.01} = 1 M_{\rm Jup}$ and $M_\star = 1.22 M_\odot$. We fix the eccentricity and argument of periapse of KOI-1474.01 to various values consistent with the light curve, then fit for the period $P_2$, the conjunction epoch $T_{0,2}$, $e_2 \cos \omega_2$, $e_2 \sin \omega_2$, and the mass $M_2$ of the perturbing body (denoted ``2'').  The fits are performed via a Levenberg-Marquardt algorithm driving a numerical integration that solves for transit times \citep{2010F}.  

\begin{deluxetable*}{cccc|ccccccc|c}
\tabletypesize{\scriptsize}
\setlength{\tabcolsep}{0.02in} 
\tablecaption{Parameter values for TTV fits.  Fixed in all fits are $M_\star = 1.22 M_\odot$,  $M_{.01} = 1 M_{\rm Jup}$, $i_{.01} = 90^\circ$, and $\Omega_{.01} = 0^\circ$. Orbital elements are Jacobian elements (the outer body's orbit referred to the center-of-mass of the star and the planet) defined at dynamical epoch BJD 2455200.\label{tab:ttvfits}}
\tablehead{
\colhead{$P_{.01}$ }  
  &  \colhead{    $T_{.01}$[BJD-2455000]      }    &  \colhead{    $e_{01}$     }    &  \colhead{    $\omega_{.01}$       }    &  \colhead{   $P_2$}  & \colhead{ $T_2$[BJD-2455000]          }    &  \colhead{     $e_2 \cos \omega_2$    }    &  \colhead{  $e_2 \sin \omega_2$    }    &  \colhead{  $M_2$ ($M_{\rm Jup}$)    } &  \colhead{  $i_2$}&  \colhead{  $\Omega_2$}   &  \colhead{  $\chi^2$} \\
  \colhead{[days]} & \colhead{[days]} &&& \colhead{[days]}&  \colhead{[days]} &&&&&
}
\startdata
&&&&&&&&&&\\
  69.709474 		& 238.271516  		& 0.74  &  $90^\circ$      	& 660.7  		& 496.0 		&    -0.0092  		&  -0.1824 		& 6.66		&$90^\circ$  &$0^\circ$ &  4.65 \\
$\pm$0.001696  	&  $\pm$0.002734 	& fixed & fixed  			& $\pm$21.0 	&   $\pm$7.2  	&  $\pm$0.0105 	&  $\pm$0.0192 	  &$\pm$0.34  & fixed&fixed&\\
&&&&&&&&&&\\
\hline
&&&&&&&&&&\\
  69.721695		&  238.150714 		& 0.90  &  $180^\circ$ 	&643.8 		& 304.81 		&  0.148  		& -0.0496   		& 5.82		&$90^\circ$ &$0^\circ$   & 2.62 \\
$\pm$0.002548  	&  $\pm$0.004422     & fixed & fixed  		&$\pm$50.6  	&  $\pm$2.31  	& $\pm$ 0.059  & $\pm$0.0103 	&  $\pm$0.98   	& fixed&fixed&\\
&&&&&&&&&&\\
\hline
&&&&&&&&&&\\
 69.749706 		& 238.303853   	& 0.74   	&  $90^\circ$      &  1038.0		& 841.9  		& -0.0681  		& -0.3567  		&  24.28		&$60^\circ$ &$130^\circ$   &  0.01 \\
 $\pm$0.000499 	& $\pm$0.000672   &  fixed 	& fixed  		& $\pm$38.5   	& $\pm$21.3   	& $\pm$0.0078    	& $\pm$0.0148     	& $\pm$0.41 &  fixed&fixed& \\
 &&&&&&&&&&
\enddata
\end{deluxetable*}

Initially we consider coplanar, edge-on orbits.  This configuration is consistent with the transiting planet, and although no transit of the perturbing body has been observed, it may transit in future data or may be within a few degrees of edge-on, which would make little difference to the TTVs. We first allow all 5 parameters of the outer planet to float freely, finding the best fits at each value.  We performed two fits (Table \ref{tab:ttvfits}, rows 1-4), one with KOI-1474.01 transiting at periapse and another with it transiting at semilatus rectum. Both fits are acceptable, so we find that we cannot currently use TTVs to distinguish these possibilities. In Figure~\ref{fig:oc}, we plot the O-C variations generated by these two models. In both cases, the perturber is a giant planet on a moderately eccentric orbit with a roughly Martian orbital period. We repeated both these fits with a fixed mass of $100 M_{\rm Jup}$ for KOI-1474.01 and found, as expected, that the solutions were similar, with only a slightly larger ($\sim 20\%$) best-fit mass for the perturber.

Next we perform a fit for which the transiting planet and the perturbing body have a 124$^\circ$ mutual inclination, a possible outcome of the secular chaos HEM mechanism \citep{2011NF}. As discussed above, non-coplanar orbits allow for a more massive perturbing companion. This fit (Table \ref{tab:ttvfits}, row 5-6), featuring a 24.3 $M_{\rm Jup}$ brown dwarf companion with a one-thousand day orbital period and moderate eccentricity, is an excellent match to the observed TTVs and is plotted in Figure \ref{fig:oc}. In contrast to the coplanar fits, this model predicts deviations not only in the central transit times but in the duration of the transits (e.g. \citealt{2002M, 2012N}), due to a secular variation in the transiting planet's duration. However, the small transit duration variations predicted by this model would not be significantly detected in the current data and, depending on the impact parameter, may or may not be detectable in by the \kep extended mission. Comparing the goodness of this fit to the two coplanar ones, we see that we can neither distinguish the orbital plane of the third body, nor limit its mass to the planetary regime. 

In all three cases, we see in the integrations that, as expected, the ``jumps" in the TTVs correspond to the companion's periapse passage. In the right panel of Figure \ref{fig:oc}, we plot the TTVs\footnote{In plotting these extended models, we have slightly adjusted the linear ephemeris of the transiting planet to remain consistent with the data while keeping future O-C variations centered at 0. Otherwise the predicted differences between the three different models appear misleadingly large.}  into the future. Additional transits in the Q7-Q12 data scheduled for future public release and through the \kep extended mission may allow us to distinguish among them, as well as the many other possible models among which we cannot distinguish currently. We have used the Bulirsch Stoer integrator in {\tt Mercury} \citep{1999C} to confirm that all three fits described here are dynamically stable over 10 Myr, with no planet-planet scattering occuring during this interval. The fits do not rule out past planet-planet scattering: in the context of HEM, the bodies could have undergone scattering in the past and subsequently stabilized as KOI-1474.01's orbit shrank through tidal dissipation. We note that the transiting planet's eccentricity undergoes secular variations and, in the case of the first two fits, the current $e_{01}$ is not the maximum and thus the planet experiences enhanced tidal dissipation during other parts of the secular cycle. We discuss this behavior further in the next section, in which we consider whether KOI-1474.01 is a failed- or proto-hot Jupiter. We defer exhaustive exploration of the parameter space of the three body model until more data are available, including additional transit times that extend the baseline to cover the perturber's subsequent periapse passage and complementary constraints on the perturber's mass, period, and eccentricity from planned radial-velocity measurements. However, the possibilities illustrated here show that pinning down the perturber's mass and orbit will likely reveal clues about the past mechanism of HEM and the future fate of KOI-1474.01. 

\section{KOI-1474.01: a proto- or failed-hot Jupiter?}
\label{sec:protofail}

KOI-1474.01 is a highly eccentric, Jupiter-sized planet being perturbed by an unseen companion, the ``smoking gun'' that may have been responsible for KOI-1474.01's HEM. The transiting planet might be either a proto-hot Jupiter that will achieve a short period, low eccentricity orbit via tidal dissipation over its host star's lifetime or a failed-hot Jupiter, too far from its star to experience significant tidal dissipation. If the planet is a failed-hot Jupiter, it is destined to spend the remainder of its host star's lifetime in the ``period valley'' \citep{2003J,2003U,2010WO}, between the region where it formed (beyond 1 AU) and the hot Jupiter region (P $<$ 10 days $\approx 0.091$~AU). 

S12 predicted the discovery of super-eccentric hot Jupiter progenitors among the \kep candidates based on the following argument. A Jupiter kicked to a small periapse via one of several proposed HEM mechanisms will enter the proto-hot Jupiter stage. Assuming that a steady flux of hot Jupiters are being spawned throughout the Galaxy, there must exist a steady-state stream of highly eccentric planets on their way to becoming the population of hot Jupiters thus far observed. The tidally-decaying Jupiters follow tracks of constant angular momentum: $\pfinal = P (1-e^2)^{3/2}$, where $P$ and $e$ are the values corresponding to any time during the circularization process.

To predict the number of highly-eccentric proto-hot Jupiters that \kep will discover, S12 used the Exoplanet Orbit Database (EOD) sample of planets with $M_p \sin i > 0.25 M_{\rm Jup}$ and $\pfinal <$ 10 days \citep[][{\tt http://www.exoplanets.org}]{2011W}. The $\pfinal$ cut-off is motivated by the excess of currently known Jupiter-mass planets on circular orbits with $P <$ 10 days. They computed the fraction of Jupiters in the ranges 3 $< \pfinal <$ 5 days and 5 $< \pfinal<$ 10 days that are moderately eccentric ($0.2 < e <0.6$). Next they multiplied these fractions by the total number of Jupiter-sized (R $>8 R_\oplus$) \kep candidates in these two $\pfinal$ ranges, yielding the predicted number of moderately eccentric \kep Jupiters. Finally, they use the \citet{1981H} tidal equations to compute the relative number of highly eccentric to moderately eccentric Jupiters at a given $\pfinal$ and predict 5-7 super eccentric Jupiters in the \kep sample with $e > 0.9$ and $P < 93$ days.

Because of the uncertainty in KOI-1474.01's eccentricity, we cannot definitively say whether it is one of the super-eccentric Jupiters predicted by S12. From our orbital fits in \S \ref{subsec:ecc}, we derive a $\pfinal$ posterior distribution of which 42$\%$ have $\pfinal < 10$ days and 19$\%$ have $\pfinal < 5$ days. Therefore, the evidence only slightly favors the interpretation that KOI-1474.01 is a failed-hot Jupiter with a $\pfinal >$ 10 days. Follow-up, high-precision radial velocity measurements may allow us to constrain KOI-1474.01's eccentricity even more tightly and confirm or rule out $e > 0.9$ and $\pfinal < 10$ days. Furthermore, the perturbing companion may cause secular variations in KOI-1474.01's eccentricity (\S \ref{subsec:ttvs}), boosting the tidal circularization rate during intervals of higher eccentricity; additional constraints on the perturber's identity may one day allow us to explore this effect.

\begin{figure*}[htbp]
\begin{centering}
\includegraphics{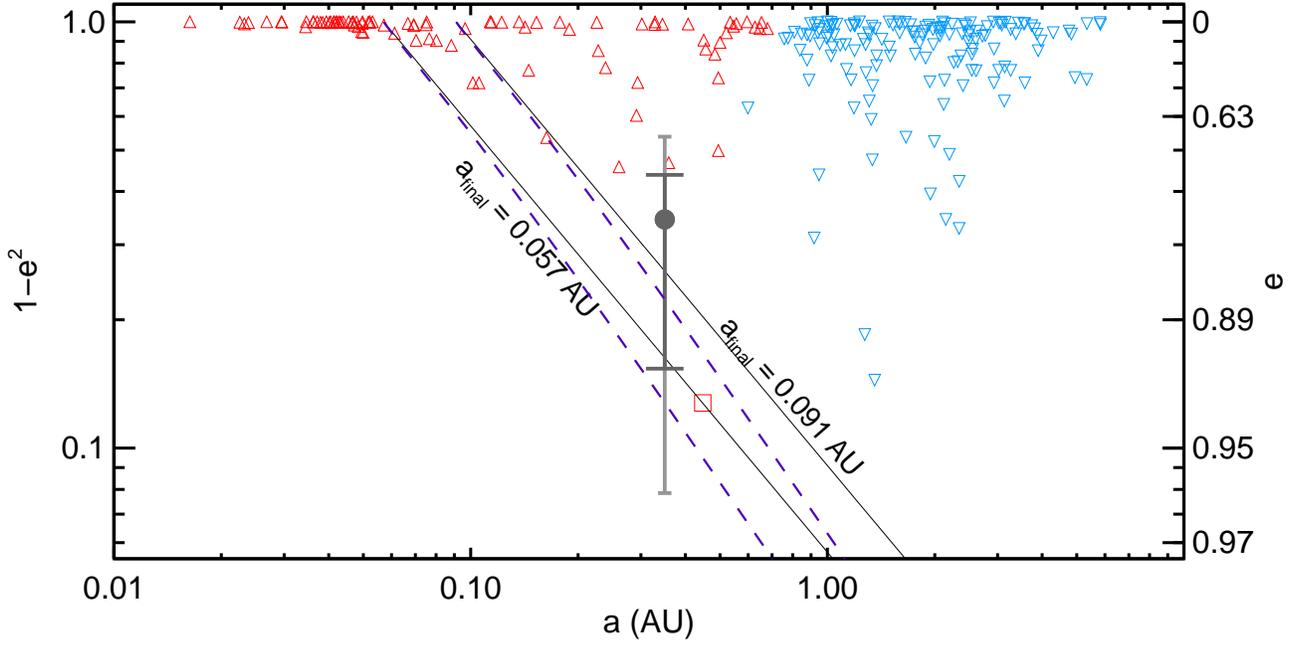}
\caption{Distribution $(1-e^2)$ vs. semimajor axis for known exoplanets from the EOD with $0.7 M_\odot < \mstar < 1.3 M_\odot$, measured eccentricities, $M_p \sin i > 0.25 M_{\rm Jup}$, and with apoapses beyond 0.9 AU (blue, downward triangles) or within 0.9 AU (red, upward triangles) \citep[][{\tt http://www.exoplanets.org}]{2011W}. The gray circle marks KOI-1474.01, with the asymmetric gray error bars representing the 1-sigma (dark gray), 2-sigma (light gray) confidence interval of KOI-1474.01's eccentricity. HD 80606 b is denoted with a red square symbol. The solid black lines are tracks of constant angular momentum corresponding to $\afinal$ = 0.057, 0.091AU; each indicates a track that a \emph{single} Jupiter follows through phase space as it undergoes tidal circularization and maintains a constant angular momentum. As it fully circularizes ($e \rightarrow 0$), a Jupiter ends up at the top of the track at $1-e^2 = 1$. The purple, dashed lines represent constant tidal circularization rates corresponding to $\acirc$ = 0.057, 0.091AU (Equation \ref{eqn:acirc}). A group of Jupiters that lie along a particular purple $\acirc$ line is undergoing tidal circularization at the same rate. \label{fig:jups} }
\end{centering}
\end{figure*}

In Figure \ref{fig:jups}, we display KOI-1474.01 (gray circle) in the context of the current sample of Jupiter-sized and Jupiter-mass planets. We plot the quantity $(1-e^2)$ vs. $a$ to allow us to overlay tracks of constant angular momentum while visually distinguishing high vs. low eccentricities. An $\afinal$ track is the path through phase space that a particular Jupiter follows during its tidal evolution; a Jupiter's current $\afinal$ defines its angular momentum and remains constant as the Jupiter undergoes tidal circularization. The solid, black lines represent tracks of angular momentum corresponding to $\afinal = 0.057, 0.091$ AU, i.e. $\pfinal = 5, 10$ days around Sun-like stars. Any Jupiter along an $\afinal$ track will stay on that track, reaching $a = \afinal$ as its $e \rightarrow 0$. The other symbols represent planets with $M_p \sin i > 0.25 M_{\rm Jup}$, $0.7 M_\odot < \mstar < 1.3 M_\odot$, and measured eccentricities from the EOD \citep{2011W}. The median of KOI-1474.01's eccentricity posterior places the planet in the period valley from $0.1~<~a~< ~1 $ AU, along with about a dozen other eccentric Jupiters. At one-sigma, KOI-1474.01 may be within (i.e. to the left of) the $\afinal < 0.057$ AU track (i.e. will end up at a semi-major axis less than 0.057 AU if it fully circularizes), like the poster-planet of high eccentricity, HD~80606~b (red square).

However, KOI-1474.01's ultimate fate is determined not only by $\pfinal$ but by its tidal circularization rate; even if the planet has $\pfinal <$ 10 days, it will not become a hot Jupiter unless it can circularize over its host star's lifetime. A hot Jupiter's tidal circularization rate depends on a combination of orbital properties and physical planetary and stellar properties. Following \citet{1998E} and \citet{2010H} --- and neglecting the effects of the planet's spin and tides raised on the star --- a tidally-circularizing planet's eccentricity e-folding time is:
\begin{equation}
\label{eqn:taue}
\frac{e}{\dot{e}} = - \frac{ a^8 (1-e^2)^{13/2} M_p}{63 \mstar^2 R_p^{10} f_e  \sigma_P}
\end{equation}
where $\sigma_P$ is the planet's internal dissipation constant and 
\begin{equation}
\label{eqn:fe}
f_e = \frac{1+\frac{45}{14}e^2+8e^4+\frac{685}{224}e^6 + \frac{255}{448} e^8 + \frac{25}{1792} e^{10}}{1 + 3e^2 + \frac{3}{8} e^4} \simeq 1+2.63 e^3
\end{equation}
Note that the tidal circularization timescale $e/\dot{e}$ depends steeply on the planet's semimajor axis and eccentricity, but only weakly on physical stellar and planetary parameters\footnote{The other parameter raised to a large power is $R_p^{10}$. Most objects with $M > 0.25 M_{\rm Jup}$ --- from Jupiters to brown dwarfs --- have $R_p \approx R_{\rm Jup}$; the $R_p^{10}$ term varies by a factor of 60 from 1 Jupiter radius to 1.5 Jupiter radius. However, in practice we find if that we normalize $a$ by $(R_p/R_{\rm Jup})^{5/4}$ for planets with known radii, Figure \ref{fig:jups} does not change significantly. The circularization rate's strong dependence on $a$ dominates, because $a$ undergoes large fractional changes throughout the hot Jupiter region, with a change in semimajor axis of 0.02 AU corresponding to an order of magnitude change in the tidal circularization timescale.}. Therefore we might expect to see a signature of tidal circularization in our $1-e^2$ vs. $a$ plot even neglecting the difference in physical properties among the planets plotted.

First imagine if all the planets underwent HEM at once and have tidally evolved for time $t$. A certain curve in $(1-e^2)$ vs. $a$ space, $\acirc (a, 1-e^2)$, represents the circularization time (Equation \ref{eqn:taue}) equal to $t$. We would expect this curve to envelope the still-eccentric Jupiter population, because all planets to the left of the curve (i.e. with $1-e^2$ less than the curve for a given semi-major axis) would have already undergone an e-folding's worth of circularization. The semi-major axis $a = \acirc$ would be the edge of the circular population we call ``hot Jupiters,'' planets for which $t$ was a sufficient amount of time to circularize. In reality, proto-hot Jupiters are being continuously spawned as new stars are born and as Jupiters undergo HEM. However, because of the steep $a$ tidal dependence -- with the tidal circularization timescale changing by an order of magnitude roughly every 0.02 AU in the hot Jupiter region --- we still expect to see an $\acirc$ boundary, corresponding to a circularization time equal to a typical stellar lifetime. To the left of this this $\acirc$ boundary would be only true proto-hot Jupiters, caught in the act of tidal circularization. With a detailed accounting for observational bias and the relatively weak effects of the planets' different physical properties, one could predict the relative number of proto-hot Jupiters on each $\acirc$ curve \citep[e.g][]{2010H}.

Solving Equation (\ref{eqn:taue}) for $(1-e^2)$, we can combine all the constants --- including the timescale $e/\dot{e}$ --- into $\acirc$ and rewrite:
\begin{equation}
\label{eqn:acirc}
(1-e^2) f_e^{-2/13} = \left(\frac{\acirc}{a}\right)^{16/13}
\end{equation}
where $\acirc$ represents the distance within which circular hot Jupiters have arrived via tidal dissipation.  For small eccentricities, the factor of $f_e$ is negligible. For large eccentricities, we can solve Equation (\ref{eqn:acirc}) numerically for $(1-e^2)$. We plot $\acirc$ curves -- along which all Jupiters have a similar tidal circularization rate -- in Figure \ref{fig:jups} as purple dashed lines. We emphasize that although the black $\afinal$ lines and purple, dashed $\acirc$ lines Figure \ref{fig:jups} are close together, their physical interpretation is different: the quantity $\acirc$ represents a proxy for the tidal circularization rate, whereas $\afinal$ is a track that an individual Jupiter follows as it undergoes tidal circularization obeying conservation of angular momentum. If the tidal evolution according to \citet{1998E} that yielded Equation \ref{eqn:acirc} is a good approximation, then $\acirc$ may be the best quantity to consider for the cut-off between proto- and failed-hot Jupiter. 

Since we see a pile-up of circular hot Jupiters and no Jupiters with $1-e^2 < 0.9$ to the left of the purple dashed line $\acirc < 0.057$ AU (P = 5 days around a Sun-like star), this may represent the timescale at which circularization happens over a fraction of a stellar lifetime. Under this interpretation, HD~80606~b's identity as a proto-hot Jupiter is not certain: it lies between $\acirc = 0.057$ AU and $\acirc = 0.091$ AU, along with several other eccentric Jupiters that have yet to circularize. Using the internal dissipation constant $\sigma_P = 3.4 \times 10^{-7} (5.9\times10^{-54})$g$^{-1}$cm$^{-2}$ derived by \citet{2010H}, the cut-off is even stricter: a Jupiter-like planet around a Sun-like star would only undergo an e-folding's worth of circularization over 10 Gyr if it had $a_{\rm circ} < 0.034$ AU. However, we note that \citet{2010H} derived the tidal dissipation constant under the assumption that proto-hot Jupiters, upon beginning their tidal circularization, have eccentricities drawn from a normal distribution with a mean $e = 0.2$ and standard deviation of 0.25. If the starting eccentricities are larger --- as assumed by S12 for proto-hot Jupiters --- a larger dissipation constant may be necessary to match the observed hot Jupiter sample. In order for a 10 Gyr e-folding time to correspond to $a_{\rm circ} = 0.057$ AU, the dissipation constant would need to be larger by a factor of 60.

The two-sigma upper limit on KOI-1474's eccentricity places the planet within $\acirc < 0.057$ AU, but the two-sigma lower limit places it well beyond this boundary. The host star's age $\tau_\star$ is currently poorly constrained (\S \ref{subsec:stellar}), and we do not know how recently the planet underwent HEM. However, if the assumptions behind the discussion above are correct, the steep dependence of the tidal circularization rate on $a$ and $e$ means that most Jupiters within $\acirc < 0.057$ AU would have circularization timescales $\ll \tau_\star$ and most Jupiters beyond $\acirc > 0.057$ would have circularization timescales $ \gg \tau_\star$. Thus the planet's fate is not sensitively dependent on either the star's age or when the planet underwent HEM; the more important quantity to pinpoint is $e$.

Finally, we note that the expected number of proto-hot Jupiters depends on the timescale for the S12 assumption of steady production. Consider the following two possibilities for the dominant HEM mechanism:
\begin{itemize}
\item HEM typically occurs on a short timescale compared to the stellar lifetime (for example, immediately as the gas disk has dissipated). Since we cannot detect planets via the transit or radial-velocity method around very young stars due to their enhanced activity, we would miss most proto-hot Jupiters, except for those in the small sliver of parameter space for which tidal circularization timescale is of order one stellar lifetime. 
\item HEM typically occurs on a timescale comparable to the stellar lifetime. In this case, we would expect to see proto-hot Jupiters at every $\acirc$, with the relative number of eccentric Jupiters (accounting for observational biases) set by the tidal circularization timescale corresponding to that $a_{\rm circ}$.
\end{itemize}

The timescale of HEM depends on which HEM mechanism is at play and on the typical initial architectures of planetary systems (e.g. for the planet-planet scattering mechanism, how tightly packed the initial configuration is). Therefore, the discovery of definitive proto-hot Jupiters would not only reveal that HEM occurs but also constrain the details of the dominant HEM mechanism. If the highly eccentric planets we find are clustered at a single $\acirc$ --- which would correspond to a tidal circularization timescale of order the stellar lifetime --- then we would conclude that HEM usually occurs early in a planetary system's history. But if highly eccentric planets are found at a range of $\acirc$ --- including  $\acirc$ within (i.e. to the left of) which most planets have circularized --- then we would conclude that HEM typically occurs throughout a planetary system's history.

\section{Discussion and future directions}
\label{sec:discuss}

We have identified KOI-1474.01 as a highly eccentric, Jupiter-sized planet using a combination of a detailed analysis of the light curve shape and the statistical validation procedure of \citet{2012M}. This makes KOI-1474.01 the second planet or planet candidate with an eccentricity measured solely via the duration aspect of the ``photoeccentric effect,'' joining KOI-686.01 whose eccentricity we measured in Paper I. We measured one component of the angle between the stellar spin axis and the planet's orbit, finding that the degree of misalignment is not currently well-constrained. Based on the variations in KOI-1474.01's transit times, we explored the identity of a perturbing companion; we found the TTVs to be consistent with perturbations from a massive, eccentric outer companion but could not uniquely constrain the perturber's mass, period, eccentricity, and mutual inclination with the currently available data. However, the main reason the perturber's parameters are poorly constrained is that we have only witnessed perturber periapse passage; we are likely to witness another periapse passage over the timespan of the \kep mission, potentially allowing us to distinguish between possible perturbers, including a coplanar giant planet vs. a brown dwarf with a large mutual inclination.

Because of the uncertainty in KOI-1474.01's measured orbital eccentricity and possible secular variations in that eccentricity due to the perturbing companion, it is not yet clear whether KOI-1474.01 is a proto-hot Jupiter --- with a periapse close enough to its star that the planet will undergo full tidal circularization over the star's lifetime --- or a failed-hot Jupiter, just outside the reach of fast tidal circularization. However, either way, the planet's discovery adds to the growing evidence that HEM mechanisms play a major role in shaping the architecture of planetary systems. The broad eccentricity distribution of extrasolar planets \citep{2008J}, the sculpting of debris disks by planets on inclined and eccentric orbits \citep[e.g][]{1997ML,1999T,2001A,2006Q,2008L,2009CK,2011D,2012DM}, the population of free-floating planets \citep{2011SK}, and the large mutual inclinations measured in the Upsilon Andromeda system \citep{2010M} all point to a dynamically violent youth for planetary systems. But the strongest evidence for HEM comes from hot Jupiters themselves --- their existence and, in many cases, misaligned or retrograde orbits \citep[e.g][]{2009WJ,2011J,2011T}. 

As a proto- or failed-hot Jupiter, KOI-1474.01 plays the crucial role of linking hot Jupiters, which are intrinsically rare, to other planetary systems. Even though they make up only a small percentage of the planet population \citep{2010HM,2011HM,2011Y,2012MM,2012W} we focus attention on hot Jupiters because, like meteorites discovered in Antarctica, they are known to come from somewhere else, bringing with them vital information about the past. In contrast, we do not know whether planets at greater orbital distances or of smaller sizes underwent migration, or if they formed in situ \citep[e.g.][]{2009V,2012H}. Moreover, the HEM mechanisms for producing hot Jupiters --- including planet-planet scattering \citep{2011N}, the Kozai mechanism \citep{2003W,2007FT,2011NF}, dynamical relaxation \citep{2008J}, and secular chaos \citep{2011WL} --- make specific predictions for the inclination distributions of hot Jupiters, which can be probed via the Rossiter-McLaughlin effect. The existence of proto- and failed-hot Jupiters will allow us to argue that the mechanisms for producing hot Jupiters are, more generally, the mechanisms that sculpt many types of planetary systems, particularly those with giant planets within ~1 AU.

The KOI-1474 system---an inner proto- or failed-hot Jupiter with a massive, long-period companion---may be the prototype of systems of hot Jupiters with distant, massive, outer companions, including as HAT-P-13 \citep{2009BH}, HAT-P-17 (\citealt{2012HB}; a hot Saturn), and Qatar-2 \citep{2012BA}. \citet{2012BA} present a compilation of the eight other hot Jupiters with known outer companions. HD~163607 \citep{2012G} resembles KOI-1474.01 in that it harbors both an eccentric inner planet (e = 0.73, P = 75.29 days) and an outer companion (in this case, a massive outer planet); however, inner planet HD~163607~b is very likely a failed-hot Jupiter, as it has $\pfinal = 24$ days. The expanding baseline for radial-velocity measurements may reveal additional, long-period outer companions of other hot Jupiters, proto-hot Jupiters, and failed-hot Jupiters \citep{2009W}. These additional companions may have been the culprits responsible for the HEM of their inner brethren. Moreover, although \citet{2012SR} examined the transit timing variations of \kep hot Jupiters and found no evidence for \emph{nearby} massive planets, the extended \kep Mission will allow for the detection of distant companions, should they exist, through TTVs.

Through radial-velocity follow up with Keck/HIRES we will measure the mass of KOI-1474.01, tighten the measurement of its high eccentricity, place additional constraints on the outer companion, and potentially discover additional bodies in the system.  Assuming a Jupiter-like composition to estimate a mass for KOI-1474.01 of $M_p \approx M_{\rm Jup}$, host star KOI-1474 would have an radial velocity semiamplitude of $\sim 70$ m s$^{-1}$, feasible for detection using Keck/HIRES.  We will then combine the RV-measured eccentricity with the transit light curves to more tightly constrain the stellar parameters, yielding a better constraint on the planet's line-of-sight spin-orbit angle $|i-i_s|$, which is currently ambiguous due to uncertainty in the stellar radius. It may even be possible to detect the Rossiter-McLaughlin effect, which has a maximum amplitude of $\approx 50$ m/s (\citealt{2010W}, eqn. 40). Although RV measurements of such a faint star ($K_P = 13.005$) pose a challenge, \citet{2012J} have demonstrated the feasibility of following up faint \kep targets with their measurements of KOI-254, a much fainter, redder star ($K_P = 15.979$). 

KOI-1474.01 contributes to the growing sample of proto- and failed-hot Jupiters. From an estimate of the unbiased number of proto-hot Jupiters, we can determine whether HEM accounts for all the hot Jupiters observed, or whether another mechanism, such as smooth disk migration, must deliver some fraction of hot Jupiters. (See \citealt{2011MJa} for the statistical methodology necessary for such a measurement.) Transiting failed-hot Jupiters orbiting cool stars will be valuable targets for testing the obliquity hypothesis of \citet{2010WF} that hot Jupiters realign cool stars: we would expect failed-hot Jupiters - which have long tidal friction timescales --- to be misaligned around both hot and cool stars.

Designed to search for Earth twins in the habitable zones of Sun-like stars, \kep is revealing a wealth of information about the origin of the most unhabitable planets of all: hot Jupiters. \emph{Kepler's} precise photometry, combined with a loose prior on the stellar density, allow us to measure the eccentricities of transiting planets from light curves alone and to search for the highly eccentric proto- and failed-hot Jupiters we would expect from HEM but not from smooth disk migration (S12). If our basic understanding of HEM and tidal circularization is correct, KOI-1474.01 is the first of a collection of highly eccentric planets that will be discovered by \kep.

\acknowledgements
We thank the anonymous reviewer for the helpful and timely report. R.I.D. gratefully acknowledges support by the National Science Foundation Graduate Research Fellowship under grant DGE-1144152. J.A.J. acknowledges support from the Alfred P. Sloan Foundation. D.C.F. (not to be confused with the DCF) is supported by NASA Hubble Fellowship HF-51272.01. A.W.H.\ acknowledges support from NASA Origins of Solar Systems grant NNX12AJ23G. We thank Zachary Berta, Joshua Carter, Courtney Dressing, Emily Fabrycky, Jonathan Irwin, Scott Kenyon, David Kipping, Maxwell Moe, Norman Murray, Smadar Naoz, and Roberto Sanchis Ojeda for helpful comments and discussions. Special thanks to J. Zachary Gazak for helpful modifications to the TAP code.

This paper includes data collected by the \kep mission. Funding for the \kep mission is provided by the NASA Science Mission directorate. Some of the data presented in this paper were obtained from the Multimission Archive at the Space Telescope Science Institute (MAST). STScI is operated by the Association of Universities for Research in Astronomy, Inc., under NASA contract NAS5-26555. Support for MAST for non-HST data is provided by the NASA Office of Space Science via grant NNX09AF08G and by other grants and contracts.

This research has made use of the Exoplanet Orbit Database and the Exoplanet Data Explorer, as well the SIMBAD database, operated at CDS, Strasbourg, France.

\bibliography{./PEHJ2} \bibliographystyle{apj}

\end{document}